\documentclass[conference,compsoc]{IEEEtran}
\ifCLASSOPTIONcompsoc
  \usepackage[nocompress]{cite}
\else
  \usepackage{cite}
\fi
\ifCLASSINFOpdf
\else
\fi
\usepackage{url}
\usepackage{amssymb}
\usepackage{amsmath}
\usepackage{xurl} % Added to break urls links
\usepackage{hyperref}

\usepackage{xspace}
\usepackage{algorithm} % captioning
\usepackage{algpseudocode}
\usepackage{enumitem}
\usepackage{amsmath}
\usepackage{amsthm}
\usepackage{comment}
\usepackage{cryptocode}
\usepackage{mdframed}
\usepackage[nameinlink]{cleveref}  
\usepackage[inkscapelatex=false]{svg}
\usepackage{diagbox} % cross out table cell
\usepackage{multirow}
\usepackage{lipsum}
\usepackage{thmtools} % restate thm
\usepackage{thm-restate} % restate thm
\usepackage{booktabs} % for table midrule etc.
\usepackage{url} % rec by NDSS for better url support, use [hyphens] to break too-long urls in double-col format
\usepackage{tcolorbox}
\tcbuselibrary{skins, breakable}

\newtcolorbox{mybox}[1][]
{
    enhanced,
    drop shadow,
    top=1mm,
    bottom=1mm,
    left=3mm,
    right=3mm,
    colback=gray!20,
    colframe=gray!90,
    sharp corners=south,
    rounded corners,
    borderline={0.5pt}{0pt}{gray!90},
    boxrule=0.8pt,
    #1
}
\usepackage{colortbl}
\newcommand{\ourrow}{\rowcolor[gray]{0.9}}

\makeatletter
\def\thmt@innercounters{equation,algorithm}
\makeatother

\declaretheorem[name=Theorem]{thm} % Use this so Cref works with restatable theorems

\newtheorem{definition}[thm]{Definition} % definition numbers are dependent on theorem numbers
\newtheorem{lemma}{Lemma}

\crefname{lemma}{Lemma}{Lemmas}
\crefname{equation}{Eqn}{Eqns}
\crefname{thm}{Theorem}{Theorems}
\crefname{section}{Section}{Sections}
\crefname{algorithm}{Algo.}{Algos}

\crefname{app}{Appendix}{Appendices}
\crefname{prop}{Prop.}{Props.}
\crefname{paragraph}{paragraph}{paragraphs}
\Crefname{paragraph}{Paragraph}{Paragraphs}

\definecolor{darkgreen}{RGB}{0,128,0}
\newcommand{\louise}[1]{{#1}}

\newcommand{\calA}{\mathcal{A}}

\newcommand{\calC}{\mathcal{C}}
\newcommand{\calD}{\mathcal{D}}

\newcommand{\calG}{\mathcal{G}}

\newcommand{\calK}{\mathcal{K}}
\newcommand{\calL}{\mathcal{L}}
\newcommand{\calM}{\mathcal{M}}

\newcommand{\calO}{\mathcal{O}}
\newcommand{\calP}{\mathcal{P}}

\newcommand{\calR}{\mathcal{R}}
\newcommand{\calS}{\mathcal{P}}

\newcommand{\calU}{\mathcal{U}}

\newcommand{\calW}{\mathcal{X}}
\newcommand{\calX}{\mathcal{X}}
\newcommand{\calY}{\mathcal{R}}

\renewcommand{\chi}{\calX}

\newcommand{\norm}[1]{\| #1 \|}
\newcommand{\R}{\mathbb{R}}
\newcommand{\Z}{\mathbb{Z}}

\newcommand{\Hfuzz}{H^\text{fuzz}_{t, \infty}}
\newcommand{\Hmin}{H_\infty}
\newcommand{\Hminavg}{\Tilde{H}_\infty}
\newcommand{\Hhill}{H^{\text{HILL}}}

\newcommand{\dist}{\text{d}}

\newcommand{\bit}{\{0,1\}}

\newcommand{\gen}{\textsc{Gen}\xspace}
\newcommand{\rep}{\textsc{Rep}\xspace}
\newcommand{\repPartial}{\textsc{Rep}\xspace}
\newcommand{\verify}{\textsc{Verify}\xspace}
\newcommand{\accept}{\textsc{Accept}\xspace}

\newcommand{\enroll}{\textsc{Enroll}\xspace}
\newcommand{\auth}{\textsc{Auth}\xspace}
\newcommand{\decode}{\textsc{Decode}\xspace}

\newcommand{\negl}{\text{negl}\xspace}
\newcommand{\poly}{\text{poly}\xspace}
\newcommand{\sd}{\textbf{SD}\xspace}

\newcommand{\adv}{\text{Adv}\xspace}

\newcommand{\comp}{\stackrel{comp}{\approx_\epsilon}}

\newcommand{\draw}[1]{\stackrel{#1}{\gets}}

\newcommand{\bob}{\text{Bob}\xspace}

\newcommand{\attackname}{{\sc Pipe}\xspace}

\newcommand{\randomfe}{\text{Randomized \fe}\xspace}
\newcommand{\prothash}{\text{Snapper-FE}\xspace}

\newcommand{\fe}{\text{FE}\xspace}
\newcommand{\owfe}{\text{OW-FE}\xspace}

\newcommand{\facialfe}{\text{Facial-FE}\xspace}
\newcommand{\facialfeE}{\text{Facial-FE E8}\xspace}
\newcommand{\facialfeLeech}{\text{Facial-FE Leech}\xspace}
\newcommand{\facialfeLeechRand}{\text{Facial-FE Leech Rand}\xspace}

\newcommand{\w}{x}
\newcommand{\x}{x}
\newcommand{\X}{X}

\newcommand{\y}{r}
\newcommand{\Y}{Y}
\newcommand{\s}{p}

\newcommand{\Lt}{$\ell_2$\xspace}
\newcommand{\asremb}{\text{ASR$_{\ell_2}$}\xspace}
\newcommand{\asrrep}{\text{ASR$_\text{Rep}$}\xspace}

\renewcommand{\l}{\ell} % Dimension of output r
\newcommand{\m}{m} % Input dimension of prot scheme
\newcommand{\n}{n} % Output dimension of prot scheme

\newcommand{\repourl}{https://github.com/MallikaPrabhakar/Embedding-Security\xspace}

\renewcommand\paragraph[1]{\par\vspace{1 ex}\noindent\textbf{#1}\,}

\begin{document}
%
% paper title
% Titles are generally capitalized except for words such as a, an, and, as,
% at, but, by, for, in, nor, of, on, or, the, to and up, which are usually
% not capitalized unless they are the first or last word of the title.
% Linebreaks \\ can be used within to get better formatting as desired.
% Do not put math or special symbols in the title.
\title{Model Inversion meets Cryptographic Fuzzy Extractors}

\author{
  \IEEEauthorblockN{Mallika Prabhakar\textsuperscript{*}, Louise Xu\textsuperscript{*}, and Prateek Saxena}
  \IEEEauthorblockA{School of Computing, National University of Singapore \\
    \{mallikaprabhakar, louisexu\}@u.nus.edu, dcsprs@nus.edu.sg}
}

% \author{\IEEEauthorblockN{Anonymous}}

% for over three affiliations, or if they all won't fit within the width
% of the page (and note that there is less available width in this regard for
% compsoc conferences compared to traditional conferences), use this
% alternative format:
% 
%\author{\IEEEauthorblockN{Michael Shell\IEEEauthorrefmark{1},
%Homer Simpson\IEEEauthorrefmark{2},
%James Kirk\IEEEauthorrefmark{3}, 
%Montgomery Scott\IEEEauthorrefmark{3} and
%Eldon Tyrell\IEEEauthorrefmark{4}}
%\IEEEauthorblockA{\IEEEauthorrefmark{1}School of Electrical and Computer Engineering\\
%Georgia Institute of Technology,
%Atlanta, Georgia 30332--0250\\ Email: see http://www.michaelshell.org/contact.html}
%\IEEEauthorblockA{\IEEEauthorrefmark{2}Twentieth Century Fox, Springfield, USA\\
%Email: homer@thesimpsons.com}
%\IEEEauthorblockA{\IEEEauthorrefmark{3}Starfleet Academy, San Francisco, California 96678-2391\\
%Telephone: (800) 555--1212, Fax: (888) 555--1212}
%\IEEEauthorblockA{\IEEEauthorrefmark{4}Tyrell Inc., 123 Replicant Street, Los Angeles, California 90210--4321}}

% use for special paper notices
%\IEEEspecialpapernotice{(Invited Paper)}

% make the title area
\maketitle
\begin{NoHyper} % Prevent hyperref warnings with empty links
\renewcommand\thefootnote{\textsuperscript{*}}
 \footnotetext{Contributed equally to this work.}
\end{NoHyper}
\begingroup\renewcommand\thefootnote{*}
\endgroup
% As a general rule, do not put math, special symbols or citations
% in the abstract
\begin{abstract}
Model inversion attacks pose an open challenge to privacy-sensitive applications that use machine learning (ML) models. For example, face authentication systems use modern ML models to compute embedding vectors from face images of the enrolled users and store them. If leaked, inversion attacks can accurately reconstruct user faces from the leaked vectors. A fuzzy extractor (FE) is a cryptographic primitive with properties that can help defend against model inversion, offering attack-agnostic security without requiring any re-training of the ML model it protects. 

To date, no systematic cryptanalysis of existing FE schemes that tolerate $\ell_2$ noise, as needed in modern ML-based face recognition systems, has been conducted. We perform in-depth security analysis of existing $\ell_2$-FE schemes showing that they offer weak security. We also show end-to-end inversion attacks that achieve high success rates in recovering original faces that are meant to be protected by FE schemes. We then offer a simple but new candidate scheme and prove its security under stated assumptions on the input distribution. Our construction mitigates all evaluated inversion attacks, and offers practical runtime
and usable accuracy for use in commodity ML-based face authentication. 
\end{abstract}

% no keywords

% For peer review papers, you can put extra information on the cover
% page as needed:
% \ifCLASSOPTIONpeerreview
% \begin{center} \bfseries EDICS Category: 3-BBND \end{center}
% \fi
%
% For peerreview papers, this IEEEtran command inserts a page break and
% creates the second title. It will be ignored for other modes.
\IEEEpeerreviewmaketitle

\newcommand{\tableLegend}{
\begin{table}[t]
\centering
\caption{Symbols used in this paper with definitions}
\label{tab:symbol-legend}
\begin{tabular}{cl}
\toprule
\textbf{Symbol} & \textbf{Definition} \\ 
\midrule
$\calM$ & {Target (feature extractor) model} \\ \hline
$\calM_{prot}$ & {Target model with post-processing protection} \\ \hline
$I$    & Input image obtained during enrollment      \\ \hline
$I'$    & Input image obtained during authentication \\ \hline
$\x$    & Input vector computed during enrollment      \\ \hline
$\x'$    & Input vector computed during authentication \\ \hline
$\x^{*}$ & {Surrogate vector recovered from $\y$}   \\ \hline
$\y$ & Output protected vector obtained from $\x$ (persistent)    \\ \hline
$\y'$    & Output protected vector obtained from $\x'$      \\ \hline
$\s$ & Output helper vector (persistent)\\ \hline
$\m$    & Size of unprotected vector $\x$      \\ \hline
$\l$ &  Size of protected vector $\y$    \\ \hline
$\n$ &  Size of ephemeral vector $b$ in~\Cref{algo:l2fe-hash} \\ \hline%
$t$ & Closeness $\ell_2$-threshold between $\x$ and $\x'$     \\ \hline
$\lambda$ & Ideal primitive security parameter
\\ \hline
$\calU_\ell$ & Uniform distribution over $\bit^\ell$     \\
\bottomrule
\end{tabular}
\end{table}
}
\newcommand{\tablePIPECross}{
\begin{table*}
\centering
\caption{ASR (in $\%$) of \attackname against different protection schemes for different datasets. \asremb is verification on output of $\calM$ while \asrrep is verification on $\calM_{prot}$. The column Guess is the ASR (mean $\pm$ std)$\%$ of a random guessing adversary.}
\label{tab:pipe-cross}
\begin{tabular}{l|ccc|ccc|ccc}
\toprule
Datasets                 & \multicolumn{3}{c|}{CelebA}                          & \multicolumn{3}{c|}{CASIA-WebFace}                  & \multicolumn{3}{c}{LFW}                              \\ \cmidrule{2-10}
Prot scheme      & \asremb & \asrrep & Guess                         & \asremb & \asrrep & Guess                         & \asremb & \asrrep & Guess                         \\ \midrule
Unprotected      & 100      & 100       & 0.9 $\pm$ 0.5  & 100      & 100       & 1.1 $\pm$ 0.7  & 99.9      & 99.9      & 1.5 $\pm$ 1.3  \\
\facialfeE      & 99.8     & 99.6      & 12.3 $\pm$ 5.8 & 99.3     & 99.5      & 9.8 $\pm$ 6.1 & 99.7     & 99.9      & 11.7 $\pm$ 6.0 \\
\facialfeLeech   & 65.1     & 72.9      & 3.0 $\pm$ 1.1  & 61.6     & 68.8      & 3.4 $\pm$ 1.5  & 65.1     & 72.7      & 4.0 $\pm$ 2.0  \\
\facialfeLeechRand & 66.5     & 73.8      & 3.0 $\pm$ 1.1  & 61.4     & 68.5      & 3.4 $\pm$ 1.5  & 63.7     & 71.6      & 4.0 $\pm$ 2.0  \\
\ourrow \prothash            & 0.4      & 3.7       & 4.1 $\pm$ 1.5  & 1.1      & 4.0         & 4.1 $\pm$ 1.4  & 0.6      & 3.6       & 4.4 $\pm$ 1.7  \\ \bottomrule
\end{tabular}
\end{table*}
}

\newcommand{\tableTPRFPR}{
\begin{table}
\centering
\caption{TPR, FPR (in $\%$) of different $\calM_{prot}$ schemes.}
\label{tab:tpr-fpr}
\begin{tabular}{l|cc|cc|cc}
\toprule
Datasets                  & \multicolumn{2}{c|}{CelebA} & \multicolumn{2}{c|}{CASIA} & \multicolumn{2}{c}{LFW} \\ \cmidrule(l){2-7} 
Prot scheme               & TPR          & FPR          & TPR          & FPR         & TPR        & FPR        \\ \midrule
Unprotected               & 91.1         & 0.6          & 76.9         & 1.5         & 94.5       & 1.2        \\
\facialfeE                & 92.9         & 11.2         & 79.7         & 8.9        & 95.5      & 11.9       \\
\facialfeLeech            & 92.3         & 3.3          & 80.3         & 3.7         & 94.8       & 4.5        \\
\facialfeLeechRand        & 92.3         & 3.3          & 80.3         & 3.7         & 94.8       & 4.5        \\
\randomfe                 & 73.1         & 6.4          & 48.6         & 4.3         & 82.8       & 5.0          \\ 
\ourrow \prothash                 & 83.5         & 4.0            & 67.1         & 4.9         & 89.4       & 5.4        \\ 
\bottomrule
\end{tabular}
\end{table}
}

\newcommand{\tableLPIPS}{
\begin{table}[t]
    \centering
    
    \caption{LPIPS metric scores for \attackname and prior attacks.}
    \label{tab:LPIPS}
\begin{tabular}{l|cccc}
\toprule
Prot scheme               & \attackname  & \bob   & GMI   & KEDMI \\ \midrule
Unprotected               & 0.210  & 0.285 & 0.248 & \textbf{0.203} \\
\facialfeE               & \textbf{0.205} & 0.290 & 0.245 & 0.212 \\
\facialfeLeech            & \textbf{0.212} & 0.287 & 0.251 & 0.224 \\
\facialfeLeechRand & \textbf{0.212} & 0.288 & 0.260  & 0.250  \\%
\ourrow \prothash                     & 0.232  & 0.502 & 0.342 & \textbf{0.221} \\\bottomrule
\end{tabular}
\end{table}
}

\newcommand{\tableFID}{
\begin{table}[t]
    \centering
    
    \caption{FID metric scores for \attackname and prior attacks.}
    \label{tab:FID}
\begin{tabular}{l|cccc}
\toprule
Prot scheme               & \attackname & \bob   & GMI    & KEDMI \\ \midrule
Unprotected               & \textbf{17.8} & 57.6  & 58.4  & 46.4 \\
\facialfeE               & \textbf{18.5} & 57.9  & 59.1  & 42.1 \\
\facialfeLeech            & \textbf{19.5} & 55.3  & 60.6  & 50.7 \\
\facialfeLeechRand & \textbf{19.1} & 57.7  & 62.4  & 54.3 \\
\ourrow \prothash                     & \textbf{18.5} & 179.1 & 101.6 & 58.5 \\ \bottomrule
\end{tabular}
\end{table}
}

\newcommand{\tablePrior}{
\begin{table}[t]
\centering

\caption{\asremb (in $\%$) of PIPE and prior inversion attacks against three protection scenarios for FaceNet embeddings. Inversion models trained and evaluated on CelebA.}
\label{tab:prior}

\begin{tabular}{l|cc|cc}
\toprule
Setting
& \multicolumn{2}{c|}{Open-set} 
& \multicolumn{2}{c}{Closed-set} \\ 
\cmidrule{2-5}
Prot scheme 
& \attackname & \bob  & GMI   & KEDMI  \\ 
\midrule
Unprotected               & 100  & 99.8 & 85.0 & 94.5  \\
\facialfeE               & 99.8 & 81.8 & 95.3 & 97.9  \\
\facialfeLeech            & 65.1 & 19.3 & 73.0 & 91.6 \\
\facialfeLeechRand & 66.5 & 0.8  & 4.7  & 5.5   \\
\ourrow \prothash                     & \textbf{0.4}  & \textbf{0.5}  & \textbf{3.1}  & \textbf{3.6}   \\ \bottomrule
\end{tabular}
\end{table}
}

\newcommand{\imgOverview}{
\begin{figure}[t]
    \centering
    % \includesvg[width=\linewidth]{images/fig1_fixed.svg}
    \includegraphics[width = \linewidth]{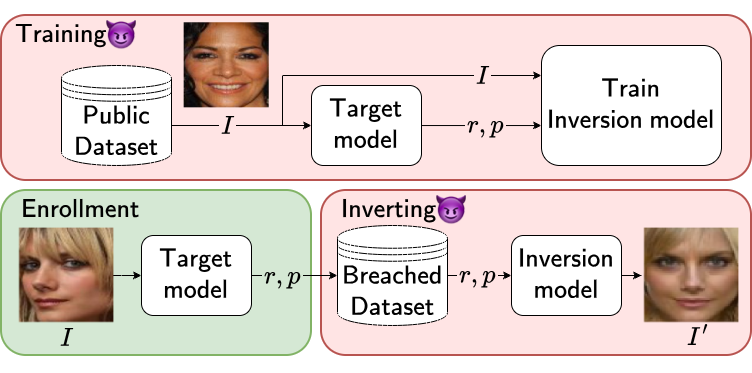}
    % \Description{Three panels. Training: images from a public dataset go to the target model, which outputs the protected embedding and helper; the attacker trains an inversion model on these image and output pairs. Enrollment: a user's face image goes through the target model, and its output is stored. Inverting: the attacker takes the stored output from a breached dataset and feeds it to the inversion model, which outputs a face resembling the user.}
    \caption{{An overview of the attack pipeline. The attacker first trains an inversion model using images from a public dataset and their protected embeddings obtained by querying the target model. At inversion time, the attacker attempts to invert a protected embedding $\y$ and helper $\s$ extracted from the breached dataset to obtain an image $I'$ that is similar to the enrollment image $I$.}}
    \label{fig:overview}
\end{figure}
}

\newcommand{\imgFaceAuth}{
\begin{figure}[t]
    \centering
    % \includesvg[width=\linewidth]{images/Pipeline.svg}
    \includegraphics[width=\linewidth]{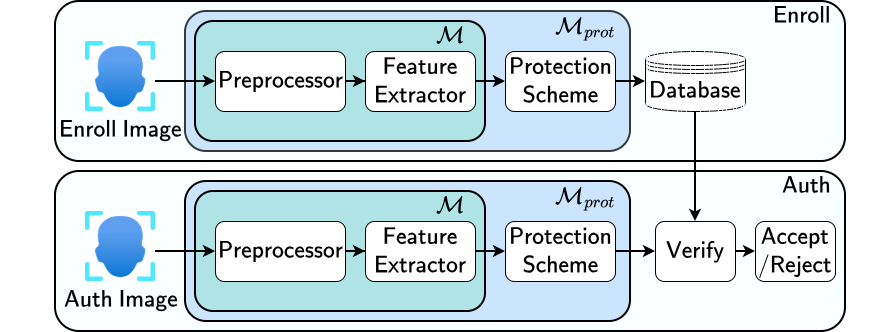}
    % \Description{Two rows, enroll and authenticate. In each row a face image passes through a preprocessor and a feature extractor (together the model M), then a protection scheme (together $M_{prot}$). At enrollment the protected output is stored in a database; at authentication the new protected output and the stored record go to a verify step that accepts or rejects.}
    \caption{A face authentication system with \enroll and \auth phases. $\calM$ represents models that output unprotected embeddings. $\calM_{prot}$ represents models with post-processing protection mechanisms that output protected embeddings.}
    \label{fig:face-auth}
\end{figure}
}

\newcommand{\imgReconstructed}{
\begin{figure}[t]
    \centering
    \includegraphics[width=\linewidth]{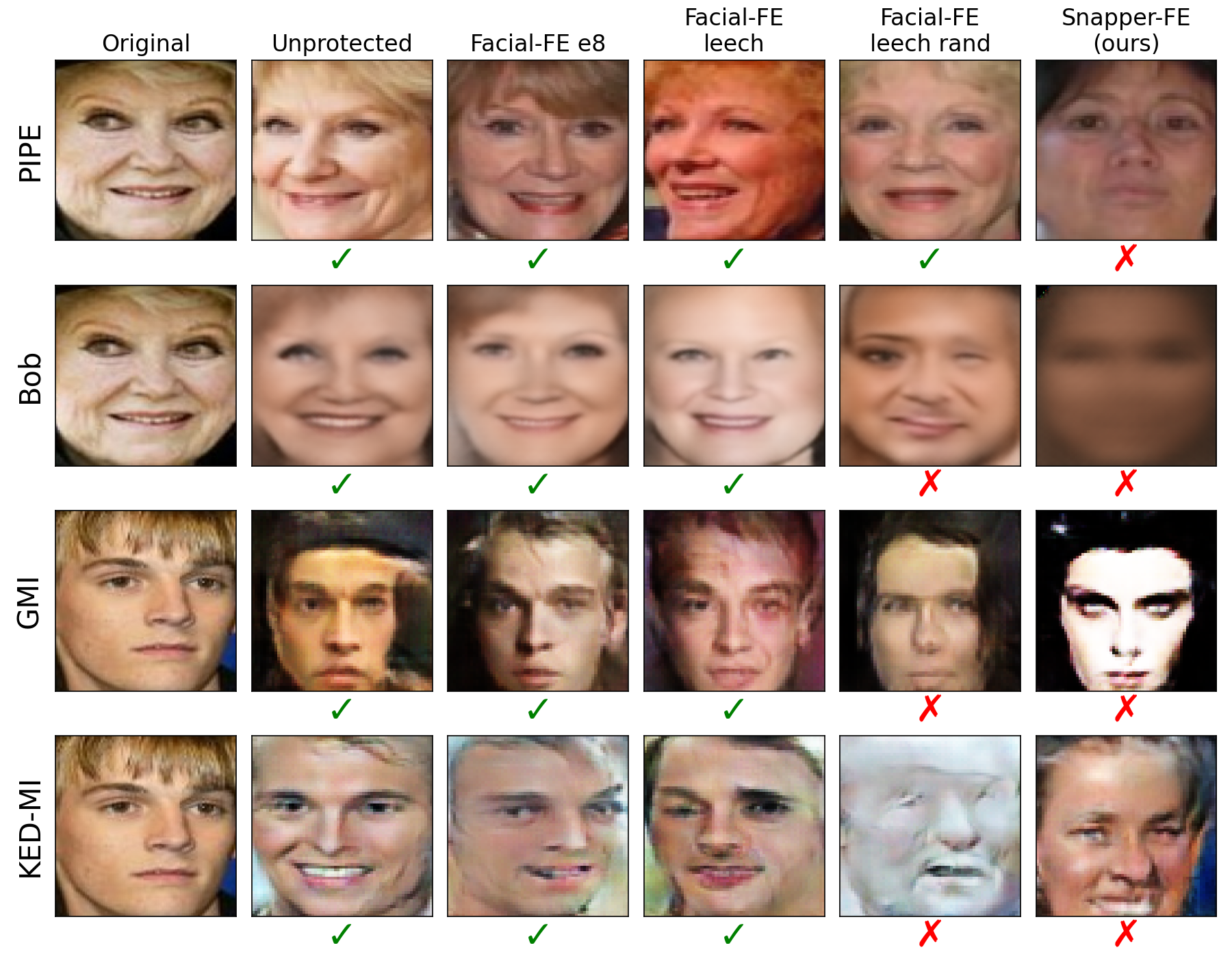}
    % \Description{Grid of face images. Rows are four attacks (PIPE, Bob, GMI, KED-MI); columns are the original face and reconstructions for unprotected embeddings, Facial-FE E8, Facial-FE Leech, Facial-FE Leech randomized, and Snapper-FE. Reconstructions of unprotected and Facial-FE embeddings resemble the original and mostly authenticate; all reconstructions from Snapper-FE fail and look unlike the original.}
    \caption{ Reconstructed images by \attackname and prior attacks against different protection schemes for FaceNet embeddings. A check/cross mark indicates authentication success/failure.
    }
    \label{fig:reconstructed-images}
\end{figure}
}

\newcommand{\imgMajorityVote}{
\begin{figure}
    \centering
    \includegraphics[width=0.9\linewidth]{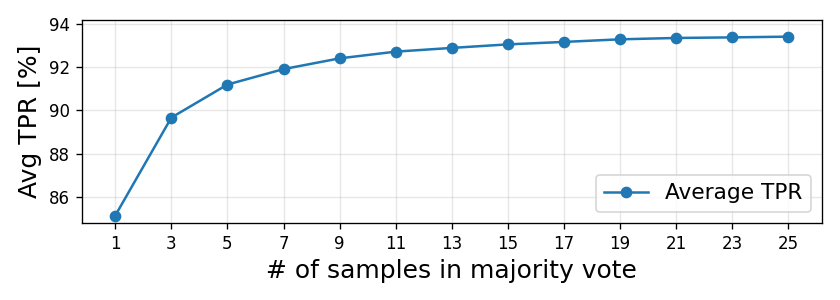}
    \includegraphics[width=0.9\linewidth]{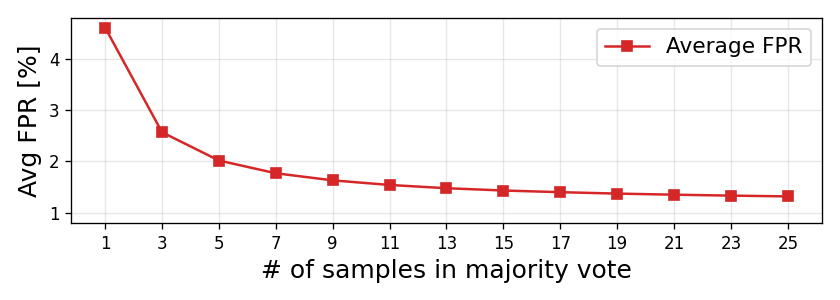}
    % \Description{Two line plots against the number of samples in the majority vote, from 1 to 25. Average TPR rises from about 85 percent with one sample to above 92 percent with nine samples and levels off near 93 percent. Average FPR falls as the number of samples grows.}
    \caption{TPR and FPR (in $\%$) of \prothash on 1000 CelebA identities for increasing samples for majority vote.}
    \label{fig:majority-vote}
\end{figure}
}

\newcommand{\imgPipeCkptASR}{
\begin{figure}
    \centering
    \includegraphics[width=\linewidth]{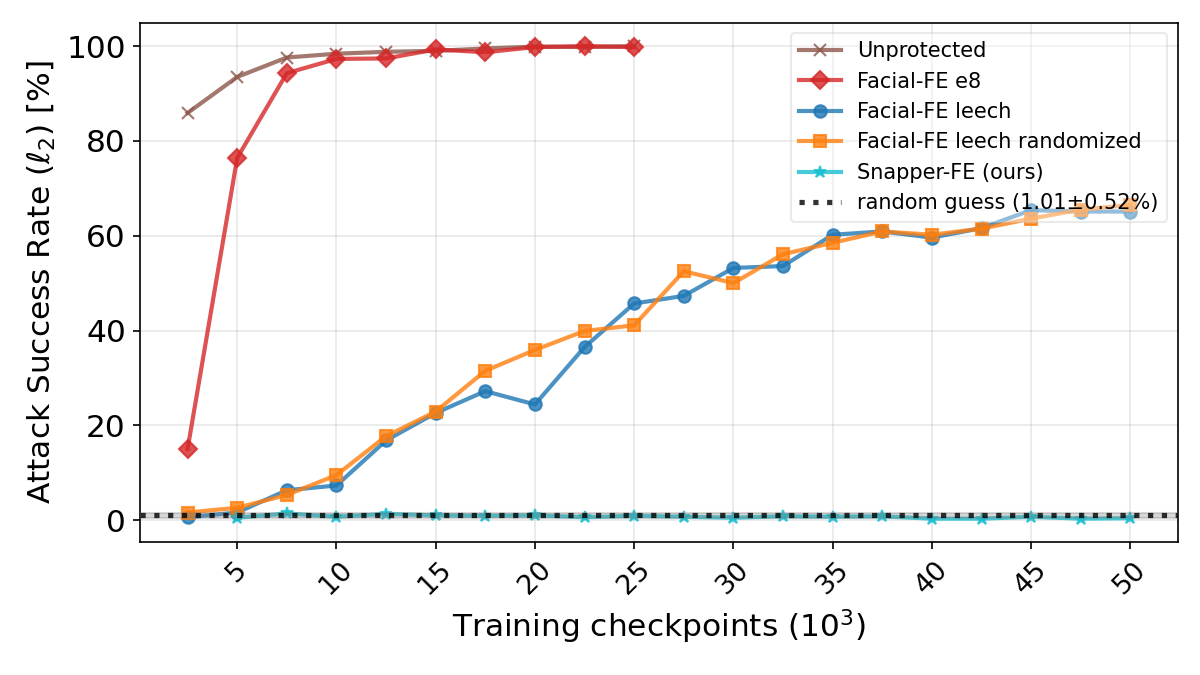}
    \includegraphics[width=\linewidth]{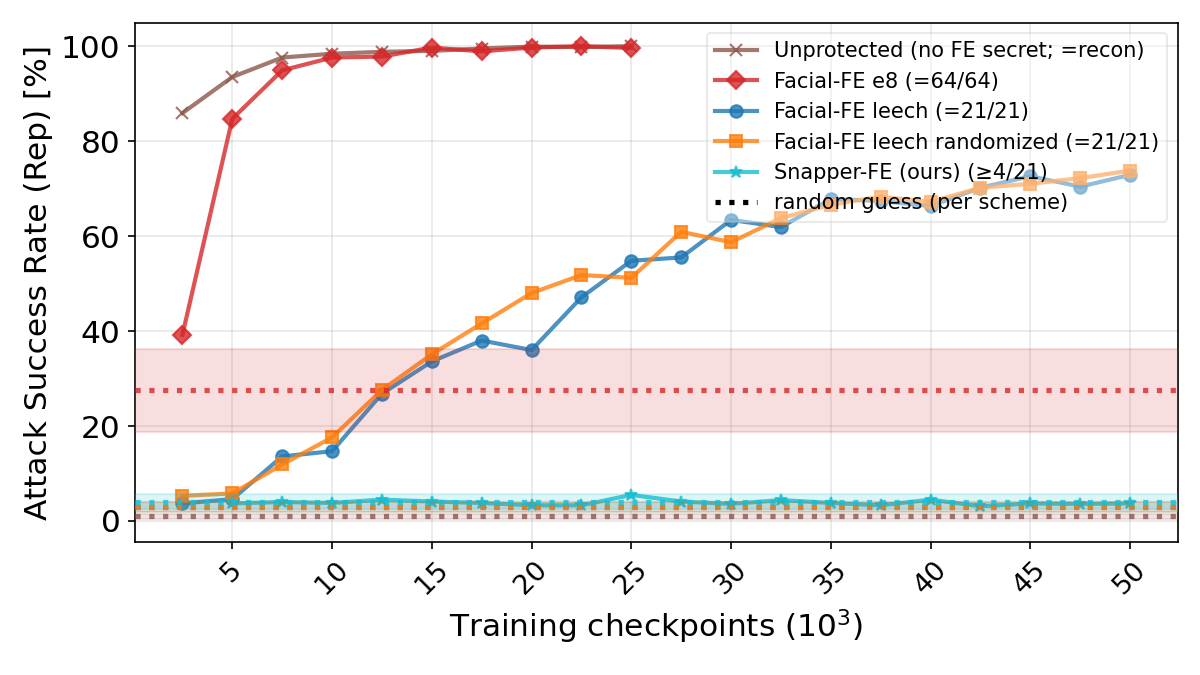}
    % \Description{Two line plots of attack success rate against training checkpoint, up to 50 thousand steps. Against unprotected embeddings and Facial-FE E8 the success rate reaches nearly 100 percent within about 10 thousand steps. Against Facial-FE Leech and its randomized variant it grows steadily, to about 65 percent under the l2 criterion. Against Snapper-FE it stays at the level of a random-guessing adversary, drawn as a dotted line with a shaded band, at every checkpoint.}
    \caption{\asremb and \asrrep of \attackname against different $\calM_{prot}$ on the CelebA dataset at different checkpoints. Dotted line and shaded region of same color indicates the  advantage (mean $\pm$ std) of a random guessing adversary.}
    \label{fig:pipe-ckpt-asr}
\end{figure}
}

\newcommand{\algoEnrollAuth}{
\begin{algorithm}[t]
\begin{algorithmic}[1]
    
    \Procedure{Enroll}{$I, \calM$}
    \State $\x \gets \calM(I)$  \Comment{$\x \in \R^m$}
    \State $(\y, \s) \gets \gen(\x)$  
    \Comment{$\y \in \R^{\ell}$}
    \State Persistently store $(\y, \s)$ in Database
    \EndProcedure
\end{algorithmic}
\begin{algorithmic}[1]
    
    \Procedure{Auth}{$I', \calM$}
    \State $\x' \gets \calM(I')$ \Comment{$\x' \in \R^m$}
    \State Load $(\y,\s)$ from Database
    \State $\y' \gets \rep(\x', \s)$
    \State Return $\verify(\y', \y)$ 
        \EndProcedure
\end{algorithmic}
\caption{ Enrollment and Authentication.}
\label{algo:enroll-auth}
\end{algorithm}
}

\newcommand{\algoFacialFEGenRep}{
\begin{algorithm}[t]
\begin{algorithmic}[1]
    
    \Procedure{Gen}{}
    \State Input: $\x \gets\calX$  
    \Comment{$\x \in \R^m$}
    \State $\x_{pp} \gets \texttt{Preprocess}(\x)$ 
    \Comment{$\x \in \Z_q^m$}
    \State $k\gets \bit^{\lambda-m}$
    \State $c \gets \texttt{Decode}(\x_{pp})$
    \Comment{$c\in\Z^m$}
    \State $s \gets c-\x_{pp}$
    \State $p \gets (s, k)$ and $r \gets H_k(\x_{pp}, s)$
    \Comment{$p\in\Z^\lambda$}
    \State \Return$(r, p)$ \Comment{$r, p$ are stored persistently}
    \EndProcedure
\end{algorithmic}
\begin{algorithmic}[1]
    
    \Procedure{Rep}{}
    \State Input: $(\x', p)$ 
    \State $\x'_{pp} \gets \texttt{Preprocess}(\x')$ 
    \State Parse $p$ as $(s,k)$
    \State $c' \gets \x'_{pp} + s$
    \State $c^* \gets \texttt{Decode}(c')$
    \State $\Tilde{\x} \gets c^*-s$
    \State \Return $H_k(\Tilde{\x}, s)$
    \EndProcedure
\end{algorithmic}
\caption{ \facialfe construction from~\cite{2021:facial-fe} Algo. 1, 2, 5, 6.}
\label{algo:facial-fe}
\end{algorithm}
}

\newcommand{\algoLtFEGenRep}{
\begin{algorithm}[t]
\begin{algorithmic}[1]
    
    \Procedure{Gen}{}
    \State Input: $\x \draw{\calX} \X$  
    \Comment{$\x \in \Z_q^m$}
    \State $A \gets \Z_q^{m \times n}$, $b \gets \Z_q^n$, $k \gets \bit^\lambda$
    \State $s \gets (A, Ab+\x)$
    \State $p \gets (s, k)$ and $r \gets H_k(b)$
    \State \Return$(r, p)$
    \EndProcedure
\end{algorithmic}
\begin{algorithmic}[1]
    
    \Procedure{Rep}{}
    \State Input: $(\x', p)$ 
    \State Parse $p$ as $(A,c, k)$
    \State $\beta \gets c- \x'$
    \Comment{$\beta \gets Ab+(\x-\x')$}
    \State $b^* \gets \texttt{Decode}(A, \beta)$
    \Comment{\texttt{Decode} is BNP}
    \State \Return $H_k(b^*)$ 
    \EndProcedure
\end{algorithmic}
\caption{ \randomfe construction.}
\label{algo:l2fe-hash}
\end{algorithm}
}

\newcommand{\algoLtFEDecodeGenRep}{
\begin{algorithm}[t]
\begin{algorithmic}[1]
    
    \Procedure{Gen}{}
    \State Input: $\x \draw{\calX} \X$ 
    \State $\x_{pp} \gets \texttt{Preprocess}(\x)$
    \Comment{$\x_{pp} \in \Z_q^{m}=\Z_q^{24\cdot N_b}$}
    \State $d \gets \decode(\x_{pp})$
    \State Parse $d$ as $d_1, ..., d_{N_b}$
    \Comment{$d_i\in\Z_q^{24}$}
    \State $k_i\gets K$ for $i=1, ..., N_b$
    \State $p \gets (k_1, ..., k_{N_b})$
    \State$r \gets  (H_{k_1}(d_1), ..., H_{k_{N_b}}(d_{N_b}))$ 
    \State \Return$(r, p)$
    \EndProcedure
\end{algorithmic}
\begin{algorithmic}[1]
    
    \Procedure{Rep}{}
    \State Input: $(\x', p)$ 
    \State Parse $p$ as $(k_1, ..., k_{N_b})$
    \State $\x'_{pp} \gets \texttt{Preprocess}(\x')$
    \Comment{$\x'_{pp} \in \Z_q^{m}=\Z_q^{24\cdot N_b}$}
    \State $d^* \gets \texttt{Decode}(\x_{pp}')$
    \Comment{$d^*_i\in\Z_q^{24}$}
    \State Parse $d^*$ as $d^*_1, ..., d^*_{N_b}$
    \State \Return $H_{k_1}(d^*_1), ..., H_{k_{N_b}}(d^*_{N_b})$ 
    \EndProcedure
\end{algorithmic}
\begin{algorithmic}[1]
    
    \Procedure{Verify}{}
    \State Input: $(\y, \y')$ 
    \State Parse $r$ as $(r_1,\ldots,r_{N_b})$ and $r'$ as $(r'_1,\ldots,r'_{N_b})$
    \State 
    $\begin{aligned}
N_{match} \gets \sum_{i=1}^{N_b} 
\mathbf{1}\!\left[
\y_{i} = \y'_i
\right]
\end{aligned}$
    \State \Return \texttt{Accept} iff $N_{match} \geq N_t$ 
    \EndProcedure
\end{algorithmic}
\caption{ \prothash construction.}
\label{algo:l2fe-hash-decode}
\end{algorithm}
}

\newcommand{\algoEntropy}{
\begin{algorithm}[t]
\caption{Residual entropy under attacker $\calA$}
\label{alg:entropy-estimation}
\begin{algorithmic}[1]
\Require Sample space $\X$, $(\gen, \repPartial)$ algorithms of \fe, attacker algorithm $\calA$, number of trials $N$
\Ensure Average residual entropy $\widehat{H}_\infty^{\calA}$

\State $\mathsf{sum} \gets 0$

\For{$i = 1$ to $N$}
    \State Sample $x \gets \X$
    \State Compute $r, p \gets \gen(x)$

    \State Let $\calR'_p \gets \calA(p, \X, \repPartial)$ be the attacker's induced guessing distribution over candidates $r'$

    \State Compute
    $
         \Hmin(\calR'_p) \gets -\log \left( \max_{r'} \Pr_{\calR'_p}[r'] \right)
    $

    \State $\mathsf{sum} \gets \mathsf{sum} + \Hmin(\calR'_p)$
\EndFor

\State \Return
$
    \widehat{H}_\infty^{\calA}
    \gets
    \frac{1}{N} \cdot \mathsf{sum}
$
\end{algorithmic}
\end{algorithm}
}

\newcommand{\algoEntropyBasic}{
\begin{algorithm}[t]
\caption{Baseline Enumeration of $\calR'_p$}
\label{alg:basic-attacker}
\begin{algorithmic}[1]
\Require Public value $p$, sample space $\X$, algorithm $\repPartial$
\Ensure Empirical guessing distribution $\calR'_p$

\State $\mathsf{Hist} \gets [\,]$

\For{$x' \in X$}
    \State Compute $r' \gets \repPartial(x', p)$
    \State Append $r'$ to $\mathsf{Hist}$
\EndFor

\State Let $\calR'_p$ be the empirical distribution induced by $\mathsf{Hist}$
\State \Return $\calR'_p$
\end{algorithmic}
\end{algorithm}
}

\newcommand{\algoEntropyBasicSoft}{
\begin{algorithm}[t]
\caption{Baseline Enumeration of $\calR'_p$ with Soft-Thresholding}
\label{alg:basic-soft-attacker}
\begin{algorithmic}[1]
\Require Public value $p$, sample space $\X = \{B_1,\ldots,B_T\}$ where each batch contains samples from one identity, algorithms $\repPartial$ and $\verify$ (wrt soft threshold $N_t$)
\Ensure Empirical guessing distribution $\calR'_p$

\State $\mathsf{Hist} \gets [\,]$

\For{each batch $B \in \X$} 
    \State $\mathsf{BatchVals} \gets [\,]$ 
    \For{each sample $x' \in B$} 
        \State Compute $r' \gets \repPartial(x',p)$ 
        \State Append $r'$ to $\mathsf{BatchVals}$ 
    \EndFor 
    \State $\mathsf{Remain} \gets \mathsf{BatchVals}$ \While{$\mathsf{Remain} \neq \emptyset$} 
        \For{each $r \in \mathsf{Remain}$} 
            \State \[\louise{\mathsf{Acc}}(r) \gets \{\, r'' \in \mathsf{Remain} : \verify(r,r'') = \accept\} \] 
        \EndFor 
        \State Choose $ r^\star \gets \arg\max_{r \in \mathsf{Remain}} |\louise{\mathsf{Acc}}(r)| $
        \State $\mathsf{Hist}[r^\star] \gets \mathsf{Hist}[r^\star] + |\louise{\mathsf{Acc}}(r^\star)|$ 
        \State $\mathsf{Remain} \gets \mathsf{Remain} \setminus \louise{\mathsf{Acc}}(r^\star)$
    \EndWhile 
\EndFor
\State Let $\calR'_p$ be the empirical distribution induced by $\mathsf{Hist}$
\State \Return $\calR'_p$
\end{algorithmic}
\end{algorithm}
}

\newcommand{\algoEntropyEliminate}{
\begin{algorithm}[t]
\caption{Elimination-Based Enumeration of $\calR'_p$}
\label{alg:elimination-attacker}
\begin{algorithmic}[1]
\Require Public value $p$, sample space $\X = \{B_1,\ldots,B_T\}$ where each batch contains samples from one identity, algorithm $\repPartial$
\Ensure Empirical guessing distribution $\calR'_p$

\State $\mathsf{Hist} \gets [\,]$

\For{each batch $B \in \X$}
    \State $\mathsf{BatchVals} \gets [\,]$

    \For{each sample $x' \in B$}
        \State Compute $r' \gets \repPartial(x', p)$
        \State Append $r'$ to $\mathsf{BatchVals}$
    \EndFor

    \If{exists $r_t$ value that is the majority in $\mathsf{BatchVals}$}
        \State Append every $r_t\in\mathsf{BatchVals}$ to $\mathsf{Hist}$
    \Else
        \State Discard $\mathsf{BatchVals}$
    \EndIf
\EndFor

\State Let $\calR'_p$ be the empirical distribution induced by $\mathsf{Hist}$
\State \Return $\calR'_p$
\end{algorithmic}
\end{algorithm}
}

\newcommand{\algoEntropyEliminateSoft}{
\begin{algorithm}[t]
\caption{Elimination-Based Enumeration of $\calR'_p$ with Soft-Thresholding}
\label{alg:elimination-soft-attacker}
\begin{algorithmic}[1]
\Require Public value $p$, sample space $\X = \{B_1,\ldots,B_T\}$ where each batch contains samples from one identity, algorithms $\repPartial$ and $\verify$ (wrt soft threshold $N_t$)
\Ensure Empirical guessing distribution $\calR'_p$

\State $\mathsf{Hist} \gets [\;]$ 

\For{each batch $B \in \X$}
    \State $\mathsf{BatchVals} \gets [\,]$

    \For{each sample $x' \in B$}
        \State Compute $r' \gets \repPartial(x',p)$
        \State Append $r'$ to $\mathsf{BatchVals}$
    \EndFor

    \For{each $r \in \mathsf{BatchVals}$}
        \State \[ \louise{\mathsf{Acc}}(r) \gets \{\, r'' \in \mathsf{BatchVals} : \verify(r,r'') = \accept\} \] 
    \EndFor

    \State Choose
    $
        r^\star
        \gets
        \arg\max_{r \in \mathsf{BatchVals}}
        |\louise{\mathsf{Acc}}(r)|
    $

    \If{$|\louise{\mathsf{Acc}}(r^\star)| \mathrel{\louise{\geq}} |\mathsf{BatchVals}|/2$}
        \State $\mathsf{Hist}[r^\star] \gets \mathsf{Hist}[r^\star] + |\louise{\mathsf{Acc}}(r^\star)|$
    \Else
        \State Discard the entire batch $B$
    \EndIf
\EndFor

\State Let $\calR'_p$ be the empirical distribution induced by $\mathsf{Hist}$
\State \Return $\calR'_p$
\end{algorithmic}
\end{algorithm}
}

\section{Introduction}
\label{sec:intro}

Machine learning (ML) systems are increasingly being used in security-sensitive applications, such as for authenticating users~\cite{2022:arcface,2015:facenet}, 
{establishing content ownership~\cite{pizzi2022imgcopydetect}}, CSAM detection~\cite{2021:neural-hash}, and so on. In these applications, neural networks often encode security-sensitive inputs into vector representations called {\em embedding vectors}. For example, in face authentication, pretrained neural networks generate embeddings corresponding to a user's face image~\cite{2022:arcface,2015:facenet}. These embedding vectors are stored in databases of application service providers~\cite{2023:clearview-ai} or on authentication devices~\cite{2024:biometric-wallet}.

Databases and devices storing sensitive information are frequently prone to breaches~\cite{2024:outabox-breach} and theft/loss~\cite{2024:biometric-wallet} 
respectively. This gives rise to a threat model where the attacker gains access to {\em all information persistently stored} in the database or the device, as is typical of attacks colloquially referred to as ``smash-and-grab" attacks. The breached information can be used to recover the original security-sensitive data corresponding to the embedding vectors, causing severe privacy risks. A well-known class of such attacks is called {\em model inversion}~\cite{2015:model-inversion}. Since their discovery in 2015 on deep networks~\cite{2015:model-inversion}, defenses with strong formal security are yet to be evaluated against model inversion attacks. 

Model inversion attacks allow the adversary to recover inputs, given the outputs of the ML model. The canonical example introduced by the original work of Fredrikson et al. showcases the threat to face recognition systems, where outputs of deep networks can be used to recover the input face images reasonably accurately~\cite{2015:model-inversion}. 
Over the last decade, ML models have evolved to serve tasks beyond classification, which motivated the original work of Fredrikson et al. The outputs of modern ML models are embedding vectors which can be compared in standard Euclidean ($\ell_2$) distance to other vectors. These vectors can be used to implement classification in a closed-set~\cite{2021:text-embedding-classification}, similarity testing in an open-set~\cite{2018:LPIPS,2015:facenet}, lookup in RAG databases~\cite{2024:rag-databases}, or generative tasks like text-to-image generation~\cite{2021:text-to-image}.
Recent works have shown that embedding vectors generated from today's ML models, if stored unprotected (in the clear), are highly susceptible to inversion attacks~\cite{2023:stylegan-recontruction-attack,2023:id3pm-diffusion-inversion-attack,2024:bob}.

Thus, it is important to devise a {\em generic defense for model inversion attacks} for applications using vector comparison with $\ell_2$ distance.
In this paper, we will use the classic example of face recognition ML models, but our goal is to motivate a broader defense strategy.

\paragraph{Cryptographic post-processing defenses.}
One way to build a generic defense with provable guarantees is to turn to cryptographic methods with clearly stated hardness assumptions. Specifically, one can post-process the outputs of the ML model with a special kind of cryptographic primitive\footnote{If we encrypt the embeddings instead, the encryption key must be persistently stored. This offers no security when all persistent data is leaked.} before they are stored, to create what we call {\em protected embeddings}, such that they can still be used for authentication while resisting model inversion attacks. The concept of such a primitive is known for over two decades in theory, called the {\em fuzzy extractor (FE)}~\cite{2008:fe-journal}. {Briefly, \fe derives random-looking outputs from noisy inputs, like face embeddings used for authentication.}

The first advantage of post-processing defenses, such as using FE primitives, is that they do not require re-training of the ML model. This contrasts with the complementary approach of in-training defense methods~\cite{2022:bido,2021:mid,2024:midre,2023:defense-gan}. {Secondly, their security depends on assumptions about the benign distribution of the embeddings they protect, so they work against arbitrary attack strategies}.
Prior in-training defenses offer no formal security hardness assumptions, nor do they aim for an attack-agnostic defense.

The third advantage is that they tolerate a powerful threat model analogous to that protected by password hashing in password database breaches~\cite{2019:ml-crack-password}. The ``password'' equivalent here is an embedding vector generated from an ML model. The adversary has query access
to the model and a public dataset. Everything stored persistently on the server or device is exfiltrated through a one-time compromise. The attacker's goal is to reconstruct the original face from the leaked embedding vector. 

In ML-based authentication, it is necessary to allow noisy authentication during normal operation. For example, face authentication requires the authenticating user to have an embedding vector that is {\it close} to the one expected. 
This is unlike textual passwords, which must match {\it exactly} and can be protected by using the standard cryptographic hash functions like \texttt{SHA256}.
Cryptographic \fe primitives allow matches between embedding vectors that are within a small distance, as needed for authentication with ML embeddings.

\paragraph{Theory-Practice Gaps.}
Most cryptographic \fe schemes are designed for use in biometric authentication such as iris, fingerprints, and so on~\cite{2011:survey-biocrypto, 2018:survey-biocrypto}. Such biometric comparisons require tolerating noise in Hamming distance or $\ell_1$ norm. 
Constructions with practical usability and security, even for $\ell_1$-norm based \fe, have only recently been demonstrated (ACM CCS'25)~\cite{2024:ccs-2025-l1}. 
Practical \fe schemes usable for authentication with ML embedding vectors require $\ell_2$-based FE---these are largely missing in prior work.

There are very few proposals for cryptographic \fe schemes that aim to work with $\ell_2$-norm noise correction as needed with modern ML models. {To our knowledge, only two such \fe constructions }%
have been proposed with any practical parameters~\cite{2021:facial-fe,2020:lwe-fe}. But none of the proposed schemes have been subject to careful cryptanalysis yet, and their direct evaluation against model inversion attacks has remained sorely missing. Therefore, there are no practically secure and usable \fe schemes that can be used with off-the-shelf face recognition systems prior to our work.

\paragraph{Our Work.} Existing $\ell_2$-\fe schemes are variants of the idea of ``code-offset'' schemes. We expose a security risk common to these code-offset constructions: the public (helper) value they output leaks partial information about the input, when considering embeddings generated from real-world face distributions as input. Thus, our first contribution is to \textbf{initiate cryptanalysis of $\ell_2$-norm based FE constructions}.

We further demonstrate that this security risk is not merely theoretical: in several practical settings, we devise concrete model inversion attacks to recover the privacy-sensitive face inputs. In particular, we present a powerful model inversion attack called \attackname which outperforms prior attacks in defeating all these \fe schemes. Experimentally, \attackname achieves an attack success rate of over $60\%$ against prior FE schemes, {compared to prior attacks that achieve as low as 0.8\% (see~\Cref{tab:prior})}.
This {successful attack} breaks their security claims, showing that prior \fe schemes do not enjoy cryptographically strong properties ideally desired.

\Cref{fig:overview} shows a successfully recovered face using \attackname from embeddings protected with an existing code-offset construction called \facialfe. The recovered image $I'$ is close to the original image $I$ both visually and as per image similarity metrics like LPIPS~\cite{2018:LPIPS}, reflecting the privacy-invasive nature of the attack. The normalized embeddings of these images are also close in \Lt distance, hence $I'$ could re-authenticate in a face authentication system. This gives way to a {\em critical, known security risk}: Face authentication on a large fraction of commercially available smartphones has been independently reported to be bypassed with a printed 2D photograph in 2023~\cite{which2023facerecognition}
and in 2026~\cite{burt2026spoofs}.

\imgOverview

We then propose \prothash, a simple $\ell_2$-based FE scheme. We show that under appropriately chosen parameters, our proposed scheme has acceptable runtime performance and accuracy compared with other alternatives in face recognition tasks with state-of-the-art ML models. We theoretically characterize its security with assumptions on input entropy, and empirically evaluate its resistance to existing inversion attacks and our new \attackname attack. 
Thus, our second contribution is a \textbf{secure candidate for $\ell_2$-norm based FE that is practically usable} in ML-based face authentication.

\paragraph{Summary of Contributions.} To our knowledge,
no prior work has connected empirical model inversion attack evaluation with cryptanalysis of $\ell_2$-based fuzzy extractors. We present \attackname, a new model inversion attack that reliably defeats existing $\ell_2$-based FE schemes, highlighting the lack of practical defenses with desirable cryptographic properties. By analyzing the insecurities of these code-offset FE schemes, we accentuate the importance of careful helper-data design to achieve security under the full-leakage threat model. To this end,
we present \prothash, a better FE scheme with practically usable runtime and security {against evaluated model inversion attacks}. We also give a formal security proof based on 
entropy assumptions of the benign input distribution, independent of the attack strategy. 

\paragraph{{Generalizing beyond faces.}}
We have chosen the face authentication system as a concrete application, aligning with the original work on model inversion~\cite{2015:model-inversion}. But, the threat of inversion attacks on embedding vectors extends well beyond that. As such, all our security properties, attacks, and definitions are generic---defined abstractly for models that produce similar high-dimensional embedding vectors for similar inputs. 
We define closeness with respect to \Lt distance since it is common in face authentication, but one can easily extend our definitions to any distance metric. One can readily consider modalities that rely on $\ell_2$ distance, like audio and text~\cite{2016:recon-audio, 2022:text-recon-attack} in future work. Likewise, one can consider other security applications such as image similarity checking for copyright infringement or CSAM detection~\cite{2021:neural-hash}.
                 
\section{The Problem}
\label{sec:problem}

Protecting long-term secrets, like authentication credentials, is an important goal in the event of catastrophic data breaches. We explain how this problem manifests in the context of vector embeddings used in modern ML models for face authentication, following the inspiration set in~\cite{2015:model-inversion}.

\subsection{Application: ML Face Authentication}
\imgFaceAuth

Billions of users have had password credentials leaked via data breaches, posing serious threats to security and privacy. Authentication services that store credentials (embeddings) derived from face images persistently, for example ML-based face authentication like ClearView~\cite{2023:clearview-ai}, Smart-ID~\cite{iproov-ml} and crypto wallets~\cite{2024:biometric-wallet}, face a similar risk of breach or theft~\cite{2024:outabox-breach}. 

A typical face authentication system employs two phases: enrollment (\enroll) and authentication (\auth), as depicted in ~\Cref{fig:face-auth}. During enrollment, the user provides an image $I$. An ML model $\calM$, often called the ``feature extractor", generates a representative embedding vector $\x\in \mathbb{R}^\m$ for $I$, which is stored as the user's credential in the database (e.g. indexed via username). When a user later tries to log into the system with a fresh image $I'$, an embedding $x'=\calM(I')$ is generated and the enrollment $\x$ is retrieved from the database (e.g. via username). The authentication request only succeeds when $\x'$ and $\x$ are close, say in the standard $\ell_2$ distance.

In summary, the authentication server stores a high-dimensional {\em embedding vector} $\x$ in the database. It expects that a legitimate user will have face images that $\calM$ maps to vectors close to $\x$. It is equally important for security that an illegitimate user will {\em not} be able to present an input image at the time of authentication that $\calM$ maps to vectors close to $\x$. Modern neural networks generate embedding vectors in various practical ways that satisfy both these properties, with acceptably low error rates, and thus are increasingly being used in production systems~\cite{2015:facenet,2022:arcface}.

\subsection{Inversion of Unprotected ML Embeddings}
Consider an attacker who gets unauthorized access to the database storing sensitive embedding vectors persistently. If no protection mechanism was implemented (i.e. $\calM$ outputs were stored directly in~\Cref{fig:face-auth}), then the attacker can access the ``unprotected" embedding $\x$, which retains substantial information about the original image. The attacker might then invert $\x$ to obtain an image that is close to the original.

Recent prior works have shown that ML models can be trained~\cite{2023:stylegan-recontruction-attack,2023:id3pm-diffusion-inversion-attack,2024:bob} to perform such inversion for standard face recognition ML systems ($\calM$) such as FaceNet~\cite{2015:facenet} and ArcFace~\cite{2022:arcface}. If successful, the inverted image is close enough to the original face image that it passes authentication with $\calM$. Prior works have reported attack success rates of over $90\%$. Thus, strong inversion attacks on unprotected embeddings are known, posing a serious privacy risk.

\subsection{Protecting Embeddings via Post-processing}

A good defense mechanism should protect embedding vectors derived from sensitive inputs when they are persistently stored. Ideally, the defense should not be attack-specific and should generically defeat all feasible attacks. Moreover, we are not seeking solutions for securing persistent database storage via additional trust in hardware~\cite{2016:sgx-explained} or in third parties to securely store cryptographic keys.

A promising approach to such a defense is via {post-processing} methods applied directly on embeddings generated from any trained ML model. They can be easily integrated into the face authentication pipeline by replacing $\calM$ with $\calM_{prot}$ shown in~\Cref{fig:face-auth}. We focus on \textbf{post-processing protection schemes} in the paper since they can be directly implemented on current models without re-training. Such post-processing defenses exist in prior cryptographic literature~\cite{2021:facial-fe,2020:lwe-fe}, which we visit in~\Cref{sec:partial-prot-scheme}.

Many prior defenses are not in the post-processing regime. They are methods that work during model training to reduce the amount of sensitive information held by the resulting embeddings~\cite{2022:bido, 2023:defense-gan, 2021:mid, 2024:midre}. But such mechanisms have had limited success. Recent works show that the defended model remains susceptible to prior inversion attacks, with $\geq 30\%$ users being re-identified for models that have accuracy $\geq 80\%$~\cite{2022:bido,2024:midre,2023:re-thinking}. Therefore, we leave this alternative line of mechanisms out of scope. Besides, they are complementary to post-processing mechanisms.

\subsection{Threat Model}

We assume the {\em full-leakage} threat model, much the same as what password hashing is meant to protect against. We assume that all {\em persistently stored} metadata {in the database }is leaked all at once, including both the protected embedding and any helper value that is strictly necessary for legitimate authentication to work. 
We focus on {inversion attacks on {\em protected embeddings}}, i.e., embeddings that have been post-processed before being stored.
Therefore, the adversary's goal is as follows: Given any leaked vector~($r$) computed as the output of \enroll, along with requisite helper~($p$), reconstruct an image that resembles the original image, close enough to authenticate as the enrolled user. Note that ephemeral data such as $(I, I', x')$ used during \enroll and \auth is not stored persistently and thus the attacker does not have access to it.

To keep our attack assumptions minimal, we work in the black-box threat model for the attacker. 
We adapt the threat model presented in~\cite{2015:model-inversion} and assume the attacker only has query (or API) access to the target model $\calM_{prot}$. The query exposes protected embeddings, instead of classification results as presumed in~\cite{2015:model-inversion} for generality, since ML models today are used in non-classification tasks too. Moreover, protection schemes used by the target models are publicly available algorithms known to the attacker.

We also assume that the attacker has access to a public dataset of face images, similar to that assumed in the original attack~\cite{2015:model-inversion}. The public dataset does not overlap with the leaked {embedding }database, and the attacker can query $\calM_{prot}$ on this {public }dataset to obtain corresponding embeddings. These assumptions are practical and common to prior attacks~\cite{2024:bob,2021:ked-mi,2020:gmi}. This is also analogous to how password crackers use real-world password distributions gleaned from leaked password databases~\cite{2019:ml-crack-password, 2019:password-crack-survey}.                 
\section{Security via One-Way Fuzzy Extractor}
\label{sec:fowh-informal}

Among the post-processing protection mechanisms studied in prior cryptographic literature~\cite{2011:survey-biocrypto, 2018:survey-biocrypto}, fuzzy extractors (FEs) stand out as a classical primitive that 
directly addresses the problem considered in this work.
We first identify three desirable properties for protecting noisy-source authentication, and argue that an \fe whose extracted key is derived with a one-way hash satisfies all three. Henceforth, we focus on such \fe-based protection schemes.

\algoEnrollAuth
\subsection{Desired Properties}
\label{sec:ideal-props}

To establish a principled basis for evaluating protection mechanisms for noisy-source authentication, we define three critical properties that an ideal primitive should satisfy.

The ideal primitive contains two subroutines: \gen and \rep. \gen takes input embedding $\x$ and outputs a protected embedding $\y$ with auxiliary helper $\s$.
On nearby input $\x'$, \rep should reproduce $\y$ {using the same helper $\s$}. Informally, an ideal primitive must satisfy the following $3$ properties; formal definitions are given in~\Cref{sec:formal-stuff}. 

\begin{enumerate}
    \item Correctness (Noise tolerance): For close inputs $\x$ and $\x'$, when $\x$ is enrolled, $\x'$ should authenticate successfully with high probability. In other words, for $\gen(\x) = (\y, \s)$, the reproduced embedding $\rep(\x', \s)=\y'$ is equal to $\y$ with high probability.
    \item Security (Fuzzy One-wayness): Given outputs $(\y,\s) = \gen(\x)$, finding any input $\x'$ close to $\x$ is hard. 
    \item Utility: The entropy of the output embeddings $\y$, conditioned on helper $\s$, should be comparable to that of the input embeddings $\x$.
\end{enumerate}

\paragraph{Authentication based on the ideal primitive.}
If such a primitive exists,~\Cref{algo:enroll-auth} shows how it readily integrates into a face authentication system, as suggested by~\cite{2008:fe-journal}\footnote{Other methods of integrating \fe into face authentication, such as~\cite{2017:fe-pke} using the \fe secret output $r$ to generate keys for a public-key encryption scheme, have also been proposed, though this distinction is not relevant.}. 
The \enroll function receives an enrollment image, applies \gen to the image's embedding vector $\x$, then stores the corresponding protected embedding $\y$ and auxiliary helper $\s$ in the database persistently. The \auth function receives an authentication image, and determines if it matches the enrollment identity.
In particular, it uses \rep to compute some $\y'$ from $\x'$ {and the stored $\s$}. The \verify function then compares $\y'$ with the stored $\y$, and outputs \accept only if $\y'=\y$. The user is allowed to log in when \verify returns \accept. 

\tableLegend

\paragraph{Why these properties?} The noise-tolerance~\Cref{prop:noise-tol} ensures that protected embeddings can be readily used in security applications that demand some error tolerance in the input. 
For example, in a face authentication system, $\calM$ generates unprotected embeddings close in \Lt norm when given similar face images. Then after applying the protection schemes, embeddings that are close under \Lt-based comparison should map to the same protected embedding, allowing \auth to perform comparison with high accuracy.

The security~\Cref{prop:approx-non-inv} addresses the risk of privacy leakage by ensuring that the embedding $\x$ is hard to recover, even approximately within some distance $t$, {\em even if we leak the helper $\s$ and the stored output $\y$}. {This is required because even recovering an approximately close input is a privacy issue. }So,~\Cref{prop:approx-non-inv} generically defeats inversion attacks under the full-leakage threat model---the protected embedding cryptographically hides the unprotected embedding it is generated from.

\Cref{prop:entropy} prevents trivial functions that compress inputs into a very small output domain, creating collisions. Even if the first two properties are satisfied, collisions would yield low entropy in stored outputs and {the primitive would }be susceptible to brute-force guessing. Moreover, it would render the primitive useless in applications that want to use the output to distinguish various inputs. For example, a primitive designed for face authentication should map input embeddings corresponding to different users to different protected embeddings. Hence, the output entropy required for~\Cref{prop:entropy} is application-dependent and often related to the input entropy. 
We use HILL entropy~\cite{1999:extractor-universal-hash-hill}, the standard computational notion of min-entropy, as a measure of the output entropy in~\Cref{prop:entropy}.

\subsection{Fuzzy Extractors}
\label{subsec:fe}

A fuzzy extractor (FE) is a celebrated cryptographic primitive~\cite{2008:fe-journal}. 
At a high level, an \fe derives high-entropy cryptographic keys from a noisy input source, like human biometrics, such that the secret key can be reliably reproduced later from close inputs~\cite{2008:fe-journal}. This makes \fe a natural fit for noisy-source authentication, where benign noise in the input should be tolerated without compromising security.

Although the formal \fe definition only requires that the key looks random given a helper string, deployed \fe systems typically derive the key using a one-way hash indexed by a public key~\cite{2021:facial-fe,2021:sample-then-lock,2024:ccs-2025-l1}.
We call an \fe whose key is the output of such a hash a \emph{one-way \fe} (\owfe), as defined in~\Cref{def:owfe}.
We prove in~\Cref{thm:fe-is-fowh} that it satisfies all three desired properties outlined
in~\Cref{sec:ideal-props}.

\paragraph{\fe-based Authentication.} 
An \fe scheme is defined by two functions \gen and \rep, whose interface aligns closely with the ideal primitive: \gen takes input $\x$ and outputs a secret extracted key $r$ and a public helper string $p$, \rep takes input $\x'$ and helper $p$ to produce some $r'$.  
Hence, an \fe scheme can be easily integrated into the face authentication system in~\Cref{algo:enroll-auth} by substituting the ideal primitive's subroutines with an \fe's \gen and \rep. 

\paragraph{Code-Offset Construction Paradigm.} 
An \fe scheme is typically instantiated using two cryptographic components: 
a secure sketch, which provides the error-correction property, and a randomness extractor, which ensures uniformity of the secret $r$~\cite{2008:fe-journal}. Commonly, the secure sketch is implemented using the code-offset construction, where a random codeword is selected, and the offset between the codeword and the input is the sketch value included in $p$.

\paragraph{Formal FE definition.}
The \fe primitive was first proposed by~\cite{2008:fe-journal}, and \cite{2020:fe-impossible} relaxed the definition to account for $\delta$ error in correctness. 
We adapt the definition of \fe with $\delta$ errors by~\cite{2020:fe-impossible} for a specific input distribution in~\Cref{def:fe-error}.
Let $X$ be a metric space with distance metric $\dist$, and $\calX$ be a distribution over $X$. 

\begin{definition}[Fuzzy Extractor with Error]
\label{def:fe-error}
    An ($X, \calX, \l, t, \epsilon_{\fe}$)-\fe with error $\delta$ is a pair of randomized procedures, ``generate” (\gen) and ``reproduce” (\rep). \gen on input $\w\in X$ outputs an extracted string $r\in \bit^\l$ and a helper string $p\in\bit^*$. \rep takes $\w'\in X$ and $p\in\bit^*$ as inputs, and outputs $r'\in\bit^\l$. (\gen, \rep) have the following properties: 
    \begin{enumerate}
        \item Correctness: If $\dist(\w, \w') \leq t$ and $(r, p ) = \gen(\w)$, then \\$\Pr[\rep(\x', p ) = r]\geq 1-\delta$. If $\dist(\w, \w') > t$, then no guarantee is provided about the output of Rep.  
        \item Security: For input distribution $\calX$, if $(\calR, \calP ) =\gen(\calX)$, then 
        $(\calR, \calP)\approx_{\epsilon_{\fe}} (\calU(\bit^\l), \calP)$.
    \end{enumerate}
\end{definition}

For computationally unbounded adversaries, $\approx_{\epsilon_{\fe}}$ means the statistical distance between two distributions is at most $\epsilon_{\fe}$. For computationally bounded adversaries, $\approx_{\epsilon_{\fe}}$ denotes advantage of at most $\epsilon_{\fe}$ for any distinguisher with runtime polynomial in $\l$. We abbreviate such probabilistic polynomial time adversaries as PPT adversaries. Correctness is evaluated in practice as the true positive rate (TPR), thus $\delta$ is reported as the false negative rate 1-TPR~\cite{2007:sketch-theory-practice,2019:iris-fe,2024:ccs-2025-l1}.

The security property of \fe guarantees that the helper $p$ can be released publicly without revealing any information about $r$, enabling $r$ to be used as a secret key for any downstream cryptographic applications. 
Notice that this is not the same as the security property of an ideal primitive. Consider a simple \fe scheme where the key is a rounding of the input, and the helper is an independent random string. Over a uniform input distribution, this scheme satisfies \fe security, yet the key reveals the input, violating fuzzy one-wayness in~\cref{prop:approx-non-inv}.
See~\Cref{app:fe-counterexample} for the full example.

Nevertheless, \fe constructions typically utilize a one-way hash to capture the hardness to invert that~\Cref{prop:approx-non-inv} requires, even if it was not stated in the original \fe definition. 
The code-offset construction paradigm applies a strong extractor like a universal hash with a public random seed~\cite{2008:fe-journal}, and many existing schemes are instantiated with cryptographic hashes~\cite{2021:facial-fe,2021:sample-then-lock,2024:ccs-2025-l1}.
In other words, they rely on a one-way hash to achieve security in practice.
We make this insight explicit by defining \owfe in~\Cref{def:owfe}, and show that a secure \owfe  is an ideal primitive
in \Cref{thm:fe-is-fowh}.

\begin{definition}[One-Way Hash Family{, cf.~\cite{2004:hash-basics}}]
\label{def:ow-hash}
    A function family $\{H_k\}_{k \in K}$ indexed by a public key $k$ from key distribution $K$, with outputs in $\bit^\ell$, is $\epsilon_{\mathsf{OW}}$-one-way if for every PPT adversary $\calA$,
    \[
    \Pr\nolimits_{k \gets K,\, u \gets \calU(\bit^\l)}\left[H_k(\calA(k, u)) = u\right] \leq \epsilon_{\mathsf{OW}}.
    \]
\end{definition}

{\Cref{def:ow-hash} requires that inverting the hash output is hard, corresponding to preimage resistance for hash functions~\cite{2004:hash-basics,1996:hac}.}

\begin{definition}[One-Way \fe]
\label{def:owfe}
    An ($\X, \calW, \l, t, \epsilon_{\fe}$)-\fe $(\gen, \rep)$ with error $\delta$ is an \owfe with one-way hash family $\{H_k\}_{k \in K}$ if, there exists procedures $\mathsf{Gen}_0, \mathsf{Rep}_0$, such that $\gen(\x)$ computes $(i, \mathsf{aux}) \gets \mathsf{Gen}_0(\x)$ and $k \gets K$, then outputs the key $r = H_k(i)$ and the helper $p = (\mathsf{aux}, k)$; and $\rep(\x', p)$ outputs $H_k(\mathsf{Rep}_0(\x', \mathsf{aux}))$.
\end{definition}

Here $i$ is an intermediate value that $\mathsf{Rep}_0$ should recover using a close input $\x'$ and public auxiliary information $\mathsf{aux}$. Thus, the rounding scheme explained earlier is a secure \fe but not an \owfe. 

{\Cref{thm:fe-is-fowh} states that an \owfe is an ideal primitive under the computationally bounded error model~\cite[Sec.~8]{2008:fe-journal}: Any PPT adversary that computes $\x'$ close to $\x$ from the stored $(r,p)$ fails to reproduce $r$ with probability at most $\alpha$. {This adversarial error bound is different from but related to correctness}. For example, a scheme with perfect correctness (i.e., $\delta = 0$ in~\Cref{def:fe-error}) has $\alpha = 0$.}

\begin{restatable}{thm}{ThmFEisFOWH}
    \label{thm:fe-is-fowh}
   Let $(\gen, \rep)$ be an ($\X, \calW, \l, t, \epsilon_{\fe}$)-\owfe with error $\delta$ whose hash family is $\epsilon_{\mathsf{OW}}$-one-way. For any PPT algorithm that outputs $\x'$ given $(r, p)$ such that $\Pr[\dist(\x,\x')\leq t \text{ and } \rep(\x', p) \neq r] \leq \alpha$, the scheme satisfies
    $\Pr[\dist(\x,\x')\leq t] \leq \epsilon_{\fe}+\epsilon_{\mathsf{OW}}+\alpha$. Hence, $(\gen, \rep)$ is an ($\X, \calW, \bit^{\l},$ $t, \lambda$)-ideal primitive where $\epsilon_{\fe}+\epsilon_{\mathsf{OW}}+\alpha \leq \negl(\lambda)$. 
    
\end{restatable}
\begin{proof}
See~\Cref{app:fe-is-fowh}
\end{proof}

{Therefore, we only analyze \owfe schemes, i.e. \fe schemes that utilize a one-way hash function when obtaining $r$, as defenses against model inversion attacks.}

\section{Existing \texorpdfstring{$\ell_2$}{l2}-\fe schemes}
\label{sec:partial-prot-scheme}

Since most \fe schemes were designed for the $\ell_1$ metric, they do not directly support face-authentication systems that rely on comparing vectors using $\ell_2$ distance. Out of the few existing $\ell_2$-\fe schemes~\cite{2016:fe-ucp-lattice,2010:fe-partition,2021:facial-fe,2020:lwe-fe}, we analyze three lattice-based schemes that are closest to providing concrete deployment: Facial-FE~\cite{2021:facial-fe} and its randomized variant, and \randomfe (which is closely related to the LWE-FE scheme~\cite{2020:lwe-fe}). Their security claims are dependent on theoretical assumptions that have not been subject to rigorous cryptanalysis with real-world face distributions. The remaining two schemes~\cite{2016:fe-ucp-lattice,2010:fe-partition} do not give concrete parameter selection criteria that we can empirically test them with.

We analyze the existing \facialfe scheme and propose a randomized variant. We find that prior model inversion attacks can break the original scheme but not its randomized variant (see~\Cref{tab:prior}). 
This failure of prior attacks does not necessarily imply security, and makes it worthy of careful investigation. 
We will show that when these code-offset \fe constructions are applied to natural face distributions, they leak partial information about the enrollment embedding that is exploitable.
Throughout this section, we use samples from a real-world face dataset CelebA~\cite{2015:celeba} and FaceNet~\cite{2015:facenet} as the feature extractor model $\calM$ to estimate the residual entropy.

\subsection{Breaking the Facial-FE Scheme}
\label{sec:partial-fe}

The prior \facialfe scheme uses dense and efficiently decodable lattices, such as the $E_8$ or the Leech lattice, to approximate $\ell_2$ error correction~\cite{2021:facial-fe}. It claims to provide computational security at 70\% false negative rate (FNR), 
with high entropy values estimated via extrapolation from impractical parameters.
The \facialfe construction is reproduced as-is in~\Cref{algo:facial-fe}, adapted to use symbols consistent with our work. The \texttt{Decode} sub-algorithm is the corresponding lattice's decoder, and $H_k$ denotes a cryptographic {one-way }hash with a random key $k$. The \texttt{Preprocess} function scales the input embedding appropriately to fit the underlying integer lattice;~\cite{2021:facial-fe} uses {an equivalent} alternative approach where the lattice is scaled instead of the input.

\algoFacialFEGenRep

\paragraph{\facialfe implementations.} 
The enrolled image embedding $x$ is preprocessed into an integer vector $\x_{pp}$, decoded to the closest lattice point $c := \texttt{Decode}(x_{pp})$, and the offset $s := c-\x_{pp}$ is released as part of the public helper string. The hash of the input and helper is stored as the persistent secret $r:=H_k(\x_{pp}, s)$. The public helper $p:=(s,k)$ is required to reconstruct the secret during authentication. {We use $\x$ and $\x_{pp}$ interchangeably when that preprocessing detail is unimportant}.

The \facialfe construction can be instantiated with different lattices for generating codewords, as suggested in their work. Dense lattices such as $E_8$ and Leech are favorable because their dense packing geometry provides many well-separated codewords while still allowing efficient decoding. This allows for noise tolerance without merging too many distinct inputs into the same output, hence improving the utility and security.

When we instantiate their construction with the $E_8$ lattice, the vector $x$ is partitioned into $N_b:=\dim(x)/8$ blocks to fit the $8$-dim lattice. Each block is independently decoded to $8$-dim lattice vectors using the lattice decoder. 
This blockwise decoding can also be viewed as decoding with the product lattice $E_8^{N_b}$~\cite{2021:facial-fe}. 

For their constructions based on the 24-dim Leech lattice, this blockwise decoding is less direct, since typical embedding dimensions (e.g. $m = 512$) are multiples of 8 but not of 24. Though the original \facialfe paper does not specify how this mismatch is handled, we use a fixed random projection during preprocessing to map each embedding to a dimension divisible by 24 (e.g. $504$). This transformation preserves pairwise $\ell_2$-distance by the JL-lemma~\cite{1984:JLLemma}.

\paragraph{\facialfe is correct.} 
\facialfe can satisfy the correctness property because linearity of lattices allows for correction of errors within the Voronoi cell, which is an approximation for correction of errors within some \Lt distance. The correctness intuition is clear: if $\x$ and $\x'$ are close in $\ell_2$ distance, then adding the same offset $s := c - \x_{pp}$ to $\x'$ would often result in $s+\x'$ lying in the same Voronoi region as $c$. Hence, the recovered $\Tilde{x}$ and the reproduced secret $\Tilde{r}:=H_k(\Tilde{x}, s)$ would be identical to the stored secret $r$.

\paragraph{\facialfe is not secure.} 
\facialfe does not satisfy \fe security property, because $s := \decode(x) - x$ reveals partial information about $x$. For natural face distributions, the offset of the enrolled embedding $x$ within its Voronoi cell may be correlated with the user identity. For example, a pair of embeddings $x$ and $x'$ corresponding to the same person can be close in $\ell_2$ distance, yet straddle the boundary of two lattice Voronoi cells. Consequently, they would decode to different lattice points. We show that this offset between an embedding $x$ and the codeword $c:=\decode(x)$ it decodes to reveals sufficient information about $x$ to be exploited. As a result, an attacker that exploits this offset information can reduce the residual uncertainty about $\calR$ after observing $p$.

\algoEntropy

\paragraph{Residual Entropy Estimation.}
To quantify the effect of this offset leakage, we measure the residual entropy of the key $r$ after observing the public helper $p$. As in~\Cref{alg:entropy-estimation}, given $p$ from a sampled enrollment, an attacker uses a finite face dataset (CelebA~\cite{2015:celeba}) to induce a guessing distribution $\calR_p'$ over plausible candidate keys. We report its min-entropy averaged over multiple samples. This is an empirical proxy for the conditional min-entropy $\Hmin(\calR \mid \calP)$ over the true face distribution, which is computationally infeasible to measure directly~\cite{2026:sample-then-correct}. 

Hence, the reported entropy values should not be interpreted as accurate estimates of the real-world entropy of human faces.  Rather, they only serve to illustrate the relative loss of entropy due to information leakage in the evaluated schemes.

We study attacks where the adversary has oracle access to both \gen and \rep procedures. We consider two instantiations of the attacker $\calA${: baseline and elimination-based. We detail the algorithms in~\Cref{app:entropy-est}. Briefly, the baseline attack enumerates the sample space $X$ to compute the distribution $\rep(X,p)$ empirically, whereas the stronger elimination-based attack utilizes the distinguishing behavior of $\repPartial(x',p)$ to get a lower-entropy empirical distribution. Notice that $x'$ samples from the correct identity reproduce the same codeword quite often to satisfy the correctness property. However, $x'$ samples from an incorrect identity often straddle lattice-cell boundaries after shifting by the offset, and thus reproduce different codewords. Therefore, same-identity inputs that result in different outputs can be eliminated as implausible candidates. }

\paragraph{\facialfe cryptanalysis.} 
Using the estimator illustrated above, the empirical residual entropy for \facialfe is reduced from a baseline of $3.62 \pm 0.77$ bits to $0.024 \pm 0.014$ bits for $E_8$-based construction, and $4.98 \pm 0.50$ bits to $0.06 \pm 0.05$ bits for Leech-based construction\footnote{Averaged over 10 random enrollments of $p$.}. While this drop does not
constitute an end-to-end inversion attack, it clearly shows the insecurity: The helper string carries exploitable information that significantly simplifies an attacker's task.

The initial entropy of say $3.62 \pm 0.77$ bits even without an attack may appear too low. This is not a security concern, but an artifact of the sample size available in the chosen face dataset (10k identities). The true min-entropy of faces of all humans, born in the past and future, is difficult to estimate. The point of security concern is the significant drop in entropy of the outputs of the \facialfe scheme.

\subsection{Breaking Randomized Facial-FE Scheme}
\label{sec:break-facial-fe-rand}

The \facialfe algorithm proposed by~\cite{2021:facial-fe} deviates from the classical code-offset construction given in~\cite{2008:fe-journal}, where the secret codeword $c$ was selected at random. So, we next modify the \facialfe scheme with a randomized codeword chosen from a fixed lattice. We now show that this does not by itself remove the exploitable signal. In fact, if $c$ were randomized ($c_{\text{rand}}$), the resulting $p_{\text{rand}}=(s_{\text{rand}},k)$ can be roughly reduced to the original $p=(s,k)$ as follows:
\begin{align*}
    &s_{\text{rand}} - \decode(s_{\text{rand}}) \\
    &= (c_{\text{rand}} - x) - \decode(c_{\text{rand}} - x)\\
    &\approx (c_{\text{rand}} - x) - \decode(c_{\text{rand}}) + \decode(x)\\
    &= \decode(x) - x = s
\end{align*}
While $\decode$ is not a strictly linear function, it exhibits near-linear properties, in the sense that $\decode(a-b)$ is a lattice point near $\decode(a) - \decode(b)$. 
Suppose $\epsilon_a := \decode(a) - a$ and $\epsilon_b := \decode(b) - b$. Then, 
$a - b = \decode(a) - \decode(b) + (\epsilon_a - \epsilon_b)$.
Notice that $\decode(a) - \decode(b)$ is also a lattice point. Hence $\decode(a-b)$ results in the same point if $(\epsilon_a - \epsilon_b)$ lies within the bounds of the Voronoi cell. Otherwise, it escapes to a nearby cell, as both $\epsilon_a$ and $\epsilon_b$ are short vectors. 

In summary, one can derive information similar to that leaked by the original public value $s$ from the $s_{\text{rand}}$. So randomizing the codeword does not make \facialfe secure.

\paragraph{Our Key Point.} 
The insecurity of helpers $p$ that we observe in both \facialfe variants, with and without randomization, is exploitable in any code-offset \fe scheme being applied to input embeddings that do not geometrically fit nicely with the structure of the codeword space, i.e. the underlying lattice. As long as the embeddings are not situated on top of the lattice in a way that every cluster of embeddings belonging to a single identity sits snugly in a Voronoi cell region, there will always be some identities whose embeddings straddle the Voronoi cell boundary, leading to a distinguisher offered by $\repPartial$ which is exploitable.

\subsection{Randomized Lattice \fe Scheme}
\label{subsec:randomizedfe}
\algoLtFEGenRep
\paragraph{Randomized FE} Inspired by the LWE-\fe scheme~\cite{2020:lwe-fe}, we present another \fe construction, which relies on randomly sampled lattices instead of fixed lattices such as Leech or E8. As seen in~\Cref{algo:l2fe-hash}, a random lattice codeword $Ab$ is selected by randomizing the underlying lattice induced by $A$ and randomizing the vector $b$. 
Each entry of $A$ is {sampled uniformly at random from $\mathbb{Z}_q$}. When presented a face embedding $x$ for enrollment{, also processed to be in $\mathbb{Z}_q$, \randomfe's \gen returns $p := (Ab+x, k), r := H_k(b)$. All vector and matrix operations in this scheme are performed over $\mathbb{Z}_q$, often denoted using$\mod q$}. When presented with $x',p$ at authentication time, \rep decodes $(Ab+x) - (x') \mod q$ by first using a closest vector problem (CVP) solver like Babai's Nearest Plane (BNP) Algorithm~\cite{1986:BNP} to obtain a nearby lattice point $Ab'$. We then left-multiply this lattice point by the {modular left inverse matrix} of $A$ to obtain $b'$. The hash $H_k(b')$ can then be compared with the stored secret for authentication. {Note that the inverse exists with high probability because $A$ is sampled uniformly at random from $\mathbb{Z}_q^{m\times n}$, and therefore has full column rank with high probability.} 

\paragraph{Difference from LWE-\fe.} While LWE-\fe stores half of the input $x$ as $r$, our construction \randomfe stores the hash of the secret vector $b$ as $r$. This modification aligns \randomfe with the \owfe formulation. As seen in~\Cref{thm:fe-is-fowh}, a hash helps achieve ideal-primitive security, so the scheme remains secure against inversion attacks even in the full leakage model where $r$ is leaked.
{Additionally,~\cite{2020:lwe-fe} presents a decoder for the Hamming distance, whereas we study $\ell_2$-norm correction with BNP algorithm.}

\paragraph{Security of \randomfe.} 
Since $H_k$ is a one-way hash function, knowing one instance of $H_k(b), A, Ab+x\mod q, k$ does not leak much information about $x$ to the attacker. However, obtaining repeated enrollments exposes new linear equations that enable exact recovery of $x$. Given $A^{m\times n}$, the attacker can compute the {modular} left kernel matrix $A_0^{m-n \times m}$ of $A$. Left-multiplying the equation $y = Ab+x\mod q$ with $A_0$ gives 
{$
A_0y = A_0Ab+A_0x = A_0x\mod q
$}. This equation does not uniquely determine $x$
since it has more unknowns ($m$) than equations ($m-n$).
However, these constraints can accumulate across multiple enrollments. An attacker needs only to obtain at least $\lceil \frac{m}{m-n}\rceil$ enrollments with different $A_i$s to get $m$ linearly independent equations, 
which uniquely determine $x$, making the scheme insecure. 
{As per the parameters in~\Cref{app:cryptanalysis-results}, we use $\lceil\frac{180}{180-60}\rceil = 2$ enrollments per image and recover the original $x$ exactly for all tested 1000 images in the CelebA test set.}

\paragraph{High overhead.} 
\label{sec:bnp-runtime} 
Importantly, \randomfe has high runtime costs.
$A \in \mathbb{R}^{m\times n}$ is a randomized lattice with no special structure for efficient decoding. The BNP decoder
incurs substantial overhead of $O(m^6\log^3(q))$, which is dominated by the LLL-reduction procedure~\cite{LLL82}. Even after projecting embeddings down to $180$ dimensions to make experiments tractable, authentication takes around $1$ minute per sample. 
If \gen precomputes the LLL-reduced Gram-Schmidt basis of $A$, then \rep can run in $O(m^2)$ time, reducing the authentication runtime to approximately $0.6$ ms per sample at the expense of longer enrollment time. While this might be usable in some applications, recall that the embeddings have already been projected down from 512 to 180 dimensions. If an authentication service provider wants to retain the original 512 dimensions for higher accuracy, 
the authentication overhead will be significantly higher and thus render their system unusable.

\section{The \attackname Attack: Inverting to Faces}
\label{sec:attack}

We
develop a model inversion attack called \attackname that exploits the insecurity of existing $\calM_{prot}$ candidates---specifically the \facialfe scheme and its variants.
\attackname uses conditional diffusion models to generate images from input vectors obtained from 
protected embeddings output by {$\calM_{prot}$}~\cite{2021:conditionedDDPM}. The success of \attackname serves as empirical confirmation of the insecurity
we explained previously.

\paragraph{Attack Setup.}
As
in~\Cref{sec:problem}, the attacker has black-box query access to the target model {$\calM_{prot}$} and access to a public face dataset. The attacker also knows the protection mechanism used by {$\calM_{prot}$}. The attacker obtains a leaked database of protected embedding vectors $r$ and the corresponding helper $p$, which the attacker wants to invert. No ephemeral secrets or intermediate values computed in~\Cref{algo:enroll-auth} are persistently stored or leaked to the adversary.

\paragraph{Attack algorithm.}
Given {$\calM_{prot}$}, the attacker can create a training set of public helper $p$ and secret $r$ values over the public image dataset. For each $(p,r)$, the attacker computes a surrogate vector $x^*$, and uses it as an input to an inversion model to recover an image close to the original image. 

\paragraph{Obtaining $x^*$.}
The construction of $x^*$ depends on the \fe scheme. For \facialfe variants, we construct $x^*$ from the components of the helper data $p$, excluding variables like the hash output $r$ and the hash key $k$. This is because the one-wayness of the underlying hash function prevents any PPT algorithm, including \attackname, from efficiently inverting and using $r$. For \prothash introduced in~\Cref{sec:proposed-prot-scheme}, the stored $r$ and $p$ consist only of the hash outputs and hash keys, so we concatenate them to form $x^*$ and pass it to \attackname, as there is no other usable value that contains information about $x$.

For \facialfe with the E8 lattice, the codewords $c$ are sparse, so the helper $s := c - x$ contains a lot of information about $x$ and is used as $x^*$. Similarly, for \facialfe with Leech lattice, we observe that $c-q/2$ is sparse, so
$x^* := (s +q/2)\mod q$. For \facialfe with randomization, we can suppress randomness due to $c_{rand}$ as per~\Cref{sec:break-facial-fe-rand}, and use $x^* := (s - \decode(s) +q/2)\mod q$. 

\paragraph{Underlying inversion model.}
There are many existing ML models that can be considered for use as inversion models. Most create good-quality but dissimilar images~\cite{2023:stylegan-recontruction-attack}, or poor-quality but representative images~\cite{2022:cnn-recontruction-attack,2024:bob}.
Recent works on generating ID-conditioned images create realistic and representative images~\cite{2024:arc2face, 2021:conditionedDDPM}. 
For our inversion model, we adapt the classifier-guided diffusion model by~\cite{2021:conditionedDDPM} and use surrogate vectors $\x^*$ as the guiding vector. The public face datasets are used to train the model to learn an association between the distribution of face images $I$ and surrogate embeddings $\x^*$.
Once trained, the inversion model is fixed for use in attacks.
At attack time (post-training), the inversion model generates images by conditioning on the surrogate $\x^*$.%

\paragraph{Defining Attack Success.}
\label{sec:asr-def}
We consider an inversion attack successful if the inverted image $I'$ authenticates successfully as the original image $I$, so attack success rate (ASR) is 
\[
    \asrrep = \frac{\#(\auth(I', I)=\accept)}{{\#(I, I') \text{ image pairs tested}}
    },
\]
where \auth uses the \gen and \rep procedures of \fe.

Since our primary objective is to prevent close-input reconstruction, 
this notion directly corresponds to the $\ell_2$ distance between the normalized (unprotected) embedding vectors output by {${\calM}$ } for $I$ and $I'$ being close. Hence we also measure
\[
    \asremb = \frac{\#(\norm{\calM(I) - \calM(I')}<t)}{{\#(I, I') \text{ image pairs tested}}}
\]
{where $t$ is the closeness threshold for $\ell_2$-norm comparison}.   
\subsection{Evaluation Setup}
\label{sec:experiments}

{We evaluate \attackname on three variants of \facialfe schemes introduced in~\Cref{sec:partial-prot-scheme}: \facialfe with $E_8$ lattice (\facialfe E8), \facialfe with Leech lattice (\facialfe Leech), and \facialfe with Leech lattice and randomization (\facialfe Leech Rand).
We do not evaluate against \randomfe, and instead refer to~\Cref{subsec:randomizedfe} which illustrates its insecurity and high overhead.
We also compare with \attackname results on \prothash, our secure candidate scheme introduced in ~\Cref{sec:proposed-prot-scheme}.

\paragraph{Models.}
We use FaceNet (512-dim version), a widely used feature extractor model, 
as the unprotected embedding model $\calM$. We augment $\calM$ with \fe as a post-processing layer to generate protected embeddings and refer to this model as the \textit{target model} {$\calM_{prot}$}. 

FaceNet has the best accuracy at face-verification tasks as determined by the DeepFace library~\cite{2024:serengil2024lightface} on the LFW dataset~\cite{21012:lfw}. It is pretrained on the VGG-Face2 dataset~\cite{2018:vggface2-dataset} with an Inception-ResNet-V1 backbone~\cite{2017:inceptionresnet} and is often used in {works on inversion attacks targeting unprotected embeddings}~\cite{2023:id3pm-diffusion-inversion-attack}.
Before passing the image to the feature extractor, we pre-process it using MTCNN~\cite{2017:mtcnn}. 

\paragraph{Thresholds.}
For comparing unprotected embeddings, instead of using the DeepFace threshold, which is chosen for the LFW dataset, we compute a new threshold using $1000$ same-user image pairs and $1000$ different-user image pairs over the train-split of the CelebA dataset~\cite{2015:celeba}. 
The threshold $t$ is determined using a decision tree algorithm~\cite{2021:chefboost} 
and is used to compare \Lt distance of normalized embeddings. The computed threshold $t$ is $1.1027$, with true positive rate (TPR) = $91.1\%$ and false positive rate (FPR) = $0.6\%$ on the CelebA test set.~\Cref{tab:tpr-fpr} contains the TPR and FPR values for the \rep procedure of each protection scheme, {where hyperparameters have been chosen to balance TPR and FPR\footnote{We optimize for parameters that maximize (TPR+1-FPR)/2.} as explained in~\Cref{app:cryptanalysis-results}}.

\paragraph{Training of Inversion Models.}
\attackname attack uses an embedding conditioned diffusion model adapted from the implementation by~\cite{2021:conditionedDDPM} as the inversion model. 
We use hyperparameters and image size of $64 \times 64$ as recommended in~\cite{2021:conditionedDDPM}, 
with a learning rate of $10^{-4}$, batch\_size of 64, ema\_rate of $0.9999$ and  $25000$ to $50000$ anneal\_steps depending on convergence.
All experiments are performed on NVIDIA A40 GPUs, where the \attackname inversion model takes around $10$ hours to train for $25000$ steps.

\paragraph{Datasets \& Ensuring no overlaps.}
To illustrate that our attack can work across multiple datasets and face distributions, {we perform {\em cross-dataset evaluation} across $3$ image datasets}: CelebA~\cite{2015:celeba}, CASIA-WebFace~\cite{2014:casia-webface}, LFW~\cite{21012:lfw}. 
The inversion model used in \attackname is trained only on the CelebA dataset, while the ASR is reported for all the $3$ datasets separately. Specifically, the inversion model is tested on $1000$ different users' images in all $3$ datasets, but trained only on $9177$ users' images in the CelebA dataset, different from the ones used in testing. Thus, there is {\em no overlap in user identities between the train and test set of \attackname}. 

\paragraph{Random Guessing.}
Before analyzing \attackname results at face value, we need to {measure }%
the advantage of an attacker who guesses randomly. If there are $1000$ possible identities for an embedding, one might naively expect an ASR for random guessing to be $1/1000 = 0.1\%$. Practically though, feature extractors on natural distributions have false positives and false negatives, leading to a higher ASR. To evaluate ASR of random guessing, we take 1000 images, each belonging to a different user, and measure how many other users' images authenticate as the enrolled user. We repeat this procedure for $1000$ different users for all three datasets. The ASR of random guessing is displayed in~\Cref{tab:pipe-cross} as mean $\pm$ std.

\subsection{Evaluation Results}
\label{sec:pipe-asr}

\tableTPRFPR
\paragraph{TPR and FPR} 
As shown in~\Cref{tab:tpr-fpr}, \facialfe on Leech lattice has the highest TPR and lowest FPR among all the protected schemes, making it the closest in performance to the unprotected scheme. 
\facialfe with $E_8$ has much higher FPR, indicating it has worse utility. While our scheme has slightly lower TPR and higher FPR relative to the unprotected setting, they still remain practical. In~\Cref{subsec:usability}, we show how TPR and FPR can be further improved for \prothash with majority voting.

\tablePIPECross
\paragraph{\attackname ASR}
We measure attack success using 2 closeness criteria, corresponding to \asrrep and \asremb as defined in~\Cref{sec:asr-def}. \Cref{tab:pipe-cross} shows that \attackname has over $68\%$ \asrrep and over $61\%$ \asremb against all \facialfe variants, reaching as high as $99.8\%$ \asremb for \facialfeE.~\Cref{app:ablation} visualizes the increase in \attackname's ASR for different protection schemes as the number of training steps increases.
\Cref{fig:reconstructed-images} shows an image reconstructed by \attackname for different $\calM_{prot}$. {For schemes with high ASR, the reconstructed image looks visually similar to the original image, while for schemes where \attackname has a lower ASR, the reconstructed image starts to look dissimilar, even if it passes the \Lt similarity check. All images reconstructed from \prothash outputs look dissimilar to the original image.} 

\begin{mybox}
    \attackname has $61.4\%$ to $99.8\%$ attack success rate against various \facialfe schemes. 
\end{mybox}

\paragraph{Prior Inversion Attacks.} We also measure ASR of 3 state-of-the-art inversion attacks: \bob~\cite{2024:bob}, KEDMI~\cite{2021:ked-mi}, GMI~\cite{2020:gmi}. We have deferred their setup details to~\Cref{app:prior-attacks}.

\tablePrior

\Cref{tab:prior} shows that prior inversion attacks have high ASR on unprotected embeddings as expected. They also have high ASR against \facialfe E8 and Leech schemes, further demonstrating their insecurity. However, prior attacks are unable to exploit \facialfeLeechRand, reporting very low ASRs. On the other hand, \attackname has significant ASR on all prior \fe schemes. It is worth noting that all the attacks, including \attackname, do not fare well against \prothash. 
This foreshadows how \prothash is a more secure scheme, as its outputs from \gen provide no useful signal for \attackname.

\begin{mybox}
    All evaluated inversion attacks, including \attackname, {perform no better than a random-guessing baseline against \prothash}.
\end{mybox}

\imgReconstructed
\Cref{fig:reconstructed-images} presents {some representative images for each attack}. 
Most \attackname reconstructions are accepted as the same face by the target model, which is {\em not} the case for \bob, GMI, and KEDMI.

\begin{mybox}
Attacks designed for unprotected embedding inversion are mitigated by \facialfe Leech randomized, but \attackname attack is highly effective against all prior defenses we evaluated. 
\end{mybox}
\section{A secure candidate scheme for \texorpdfstring{$\ell_2$}{l2}-norm}
\label{sec:proposed-prot-scheme}

We propose a simple $\ell_2$-based protection 
scheme called \prothash as shown in~\Cref{algo:l2fe-hash-decode}. It differs from the typical code-offset paradigm as we first decode the preprocessed input into a lattice vector, and store its hash as the secret while releasing the hash key publicly. 
This {\em decoding-first} paradigm deliberately sacrifices some correctness for security, but we argue that this is the correct trade-off to make.
Correctness could be improved in practice by using multiple samples at authentication-time, but a security leakage from public information would create an attack surface that model inversion attacks could potentially exploit. 

\algoLtFEDecodeGenRep

\paragraph{\prothash~Construction.} The set of finite lattice codewords is instantiated blockwise by repeating the underlying lattice $\calL_q$ for $N_b$ times. The \gen algorithm preprocesses input embedding $\x$, decodes it to $d:=\decode(\x_{pp})$.   
We use a block-wise hash of the lattice codeword $d$ as the secret $r$, and the hash keys are released as public information.

In our implementation, $\calL_q$ is instantiated as the Leech lattice modulo $\Z_q$ with $q=1024$ because it is the densest lattice in 24 dimensions. The Leech lattice-based construction also offers the best security level among all variants of \facialfe, as claimed by~\cite{2021:facial-fe} and supported by our \attackname results. Similar to our implementation of the Leech-based \facialfe, we use a fixed projection to transform inputs into $504$ dimensions and implement \decode per $24$-dimensional block. 
We instantiate $H_k$ with keyed BLAKE2b with $\ell' = 32$ output
bits, and analyze it in the random oracle model~\cite{BR93}, as is common for the hashes in prior \fe implementations~\cite{2019:iris-fe,2024:ccs-2025-l1,2021:sample-then-lock,2026:sample-then-correct}.

We employ a blockwise authentication procedure:
\gen and \rep compute hashes per block, and \verify accepts when at least $N_t$ blocks' hashes match. This matches the block-wise Leech-lattice implementation and improves utility. 
We use threshold $N_t=4$ out of $N_b=21$, which gives the best TPR-FPR trade-off empirically. 

\subsection{Security Analysis}

We formally prove the security of our scheme under conditional min-entropy assumptions stated in~\Cref{thm:new-scheme-sec}.
Intuitively, decoding the input into codewords removes the geometric structure exploited by attacks such as \attackname, while hashing prevents the codewords from serving as useful attack-conditioning signals.
\Cref{thm:new-scheme-sec} establishes the entropy assumptions under which the \prothash output is pseudorandom given the helper data. Then hardness of reconstructing a close input follows from \Cref{thm:snapper-fow} in~\Cref{sec:new-scheme-sec-proof}, proved under the same assumptions.

\begin{thm}
\label{thm:new-scheme-sec}
For an integer lattice $\calL$, let $\calL_q := \calL \mod q$. 
Let $\calX$ denote the input distribution over $\Z_q^m$, where $m=N_b\cdot \dim(\calL)$. Let $\calL_q^{N_b}\subseteq \Z_q^m$ denote the set of lattice codewords. Let $\calD \gets \decode(\calX) \subseteq \calL_q^{N_b}$, and $\calD_i\gets\decode(\calX_i)$ denote per-block distribution for blocks $i=1,...,N_b$. Let $h$ denote the minimum per-block conditional min-entropy, i.e. $\Hmin(\calD_i\mid \calD_1, ..., \calD_{i-1})\geq h\,\forall i$. 

For $\epsilon > 0$, let $\{H_k \colon \calL_q \to \bit^{\ell'}\}_{k\in K}$ be a universal hash family where $\ell' \leq h - 2\log(1/\epsilon)+2$. Let $\calK_1, ..., \calK_{N_b}$ be independent uniform distributions over $K$, and $\ell := N_b\ell'$.
Define $\calR_i := H_{\calK_i}(\calD_i), \calR := (\calR_1, ..., \calR_{N_b}), \calP:= (\calK_1, ..., \calK_{N_b})$.
Then
$
\sd((\calR, \calP), (\calU_{\ell}, \calP)) \leq  N_b\epsilon.
$
\end{thm}

\begin{proof}
See~\Cref{sec:new-scheme-sec-proof} for the full proof.
\end{proof}

In~\Cref{sec:new-scheme-sec-proof}, we spell out how \prothash satisfies~\Cref{prop:entropy,prop:approx-non-inv} 
by~\Cref{thm:new-scheme-sec,thm:snapper-fow} respectively.
Together with empirical accuracy measured in~\Cref{subsec:usability}, it is a secure candidate for an ideal primitive.

\textbf{Remark.} We do not assume that the input distribution $\calX$ is uniform or Gaussian, unlike prior works~\cite{2010:fe-partition,2020:lwe-fe}. Our main assumption is much weaker: the input distribution has sufficient minimum per-block conditional min-entropy after decoding.%

\paragraph{Scope and future work}
Our theorems bound the privacy risk of recovering a close input from the stored $(r,p)$. It is not affected by the threshold applied at authentication time. However,
a low threshold $N_t$ can increase impersonation risk, as evidenced by the $4\%$ FPR (\Cref{tab:tpr-fpr}). Nevertheless, this does not directly translate to a privacy risk.
We leave analysis of the threshold's effect on impersonation attacks and measurement of $h$ for real-world faces to future work as they are beyond the scope of our work.
Promising avenues to increase $N_t$ and boost the entropy $h$ are fine-tuning the embedding model, combining complementary biometric modalities~\cite{fohr2025deep,mai2015fusing}, or hashing several blocks jointly as in sample-then-lock~\cite{2021:sample-then-lock}.

\paragraph{\prothash cryptanalysis.}
{For more empirical evidence of security, we estimate residual entropy for our scheme with }the baseline and elimination-based attacker, modified to consider blockwise thresholding; see~\Cref{sec:entropy-est-soft} for details. 
We observe only a {\em negligible reduction in entropy} due to the elimination attack, from $12.50$ bits (inputs) to $12.45$ bits (outputs) over the $\sim$202k samples in the CelebA dataset, consistent with our estimation procedure for \facialfe
.

This is expected because when TPR is high,  
decoding multiple samples of $x'$ from the same identity will result in the same codeword $c'$ with high probability, and the block-wise hash results would match with each other consistently, with respect to the $N_t$ threshold. In other words, running $\rep$ with samples from the correct identity or the incorrect identity would both reproduce consistent behavior, leaving no distinguishing behavior to be exploited. This makes the elimination attack ineffective on our scheme.

\paragraph{Efficacy against model inversion attacks.}
As demonstrated in~\Cref{sec:pipe-asr},
\attackname only achieve near $0.4\%\sim1.1\%$ \asremb and $3.6\%\sim4.0\%$ \asrrep
against our scheme, and prior attacks do no better (see~\Cref{tab:pipe-cross,tab:prior}). This is comparable to the random-guessing baseline of $4.1\%\sim4.4\%$, 
as clearly evidenced by~\Cref{fig:pipe-ckpt-asr} in~\Cref{app:ablation}.  

\subsection{Usability and Runtime Costs}
\label{subsec:usability}

We also measure the empirical performance of \prothash and observe practically reasonable TPR and FPR values and runtime.

\paragraph{Usability.} It is easy to see that when two same-identity embeddings
close in $\ell_2$ distance decode to different lattice codewords, their hashes would produce two different $r$. In other words, the decoding step may negatively impact correctness. But, we empirically justify why this is an appropriate design trade-off favoring security. 

We see in ~\Cref{tab:tpr-fpr} that across three datasets, TPR reduces from $76.9\%\sim94.5\%$ to $67.1\%\sim89.4\%$ and FPR increases from $0.6\%\sim1.5\%$ to $4.0\%\sim5.4\%$ for our scheme, compared to the unprotected embedding setting. 
The reduced utility is not too severe,
as TPR and FPR can be further improved by taking a majority vote over multiple authentication 
vectors~\cite{2011:majority-vote}. As mentioned, they can also
be improved further by fine-tuning the feature extractor so embedding blocks sit nicely within the lattice.

\paragraph{Boosting accuracy with Majority voting.}
A real deployment such as~\cite{singpass} can take multiple
authentication-time measurements of a person's face and compare against the single stored protected embedding.
It can check success across all measurements and take a majority vote in the final decision. We demonstrate that this simple trick improves TPR to over $92\%$ and reduces FPR to below $2\%$ via an experiment on $1000$ random identities. This performance is comparable to the performance of the unprotected model (\Cref{app:ablation}).
The security theorems still hold since the stored information $(r,p)$ remains the same regardless of authentication behavior.

\paragraph{Runtime Cost Comparisons.} \prothash has nearly the same runtime performance as \facialfe schemes based on Leech lattices, since we use the same decoders---the dominant cost. Practically, we observe authentication time to be about $0.9$ ms per sample on the $504$-dimensional embeddings. It is comparable to the roughly $0.6$ ms per sample runtime for \randomfe after LLL-reduction and inputs projected to $180$ dimensions. Multiple authentication attempts can be computed in parallel and are not a bottleneck.
\section{Related Work}

An embedding inversion attack is a canonical example of model inversion attacks that targets model embeddings~\cite{2015:model-inversion}. 
A number of model inversion attacks on image recognition ML models have been developed, often targeting image classification outputs~\cite{2021:ked-mi,2020:gmi,2021:vmi,2023:re-thinking,2023:plgmi}. There are also inversion attacks on face authentication systems that target image embeddings~\cite{2024:bob, 2023:id3pm-diffusion-inversion-attack}. We have evaluated $3$ prior attacks representative of SOTA attack accuracy. 

{Gradient inversion attack is another type of model inversion attack that uses model gradients to infer private training data~\cite{2019:gradient-inversion}. Since this requires an adversary to have gradient information, it does not fit our black-box authentication threat model. }

\paragraph{Non-post-processing defenses.} Many defenses against model inversion modify the training process, not post-processing outputs of ML models already trained~\cite{2022:bido,2021:mid,2024:midre, 2023:defense-gan}. 
Because these defenses are highly model-specific and optimization-based, it is hard to theoretically characterize their end security guarantees.
Their recently reported effectiveness against existing black-box attacks is fairly limited, with attacks re-identifying $\geq 30\%$ of users from reconstructed images from models that have accuracy $\geq 80\%$~\cite{2024:midre,2022:bido}.  
Importantly, our work points towards defining properties of cryptographic defenses that generically work with ML models for authentication and in an attack-agnostic way.

The privacy risk from model inversion is immediate, which is our primary concern. There is also re-authentication risk, i.e., using recovered face images to impersonate the user. Liveness detectors during face authentication mitigate such risks, but a line of prior works originating with Xu et al.~\cite{2016:defeat-liveness-detect} shows attacks against such mechanisms. These techniques are complementary to our study.

\paragraph{Post-processing protection schemes.}
Most post-processing protection mechanisms for embedding vectors are some variant of multispace random projection (MRP)~\cite{2007:mrp}, which are not secure under the full-leakage threat model.
MRP variants like BioHashing~\cite{2004:biohashing} and Index-of-Max Hashing~\cite{2018:iom-hash} are also vulnerable to attacks~\cite{2005:bh-preimage-attack-withR, 2009:bh-preimage-attack-withoutR,2020:iom-hash-attack}.
In fact, many protection mechanisms that rely on a distance-preserving transformation are vulnerable to pre-image attacks~\cite{2024:rp-attack, 2019:genetic-preimage-attack}, and thus do not satisfy~\Cref{prop:approx-non-inv}. Some schemes combine MRP with fuzzy commitment for additional security, but they still do not address inversion attacks that recover approximately close inputs~\cite{2019:rp-fc, 2021:rp-fc-noninvert}.

Similarly, fuzzy hashes, or locality-sensitive hashes~\cite{1999:locality-sensitive-hash, 2014:locality-preserve-hash} aim to find linear projections that preserve local information. For example, Apple's NeuralHash trains a CNN model to learn such mappings for security applications, but it remains vulnerable to adversarial attacks that force hash collisions~\cite{2021:neural-hash, 2022:neural-hash-attack,2026:break-neuralhash-faces}.
In general, CNN models are particularly susceptible to model inversion attacks~\cite{2022:cnn-recontruction-attack}.

{A complementary approach that builds upon code-offset \fe schemes is to train an additional expander layer to ensure the embeddings align better with the Voronoi cell regions, as in~\cite{jana2022neuralfe} for fingerprint-based authentication. Our \prothash requires no additional training, while remaining compatible with such distribution fine-tuning approaches.}

\paragraph{Other Threat Models.}
Our desired security definition under the full-leakage threat model is not to be confused with post-compromise security. The latter addresses the issue of secure encryption of future messages after key leakage, whereas~\Cref{prop:approx-non-inv} addresses the issue of information leakage about past messages after key leakage~\cite{2016:post-compromise}. 

\Cref{prop:approx-non-inv} is also different from forward secrecy, a property of key-exchange protocols ensuring that compromising a long-term key does not reveal past session keys~\cite{2019:forward-secrecy}. The fundamental idea that a compromise should not lead to decrypting past messages is similar, but our setting is different from a key-exchange protocol. We additionally assume past encrypted messages, which in our context are protected embeddings, can be compromised. 

Honey encryption (HE) addresses a different threat model from ours~\cite{2014:honey-enc}. It allows for encryption of passwords with low-entropy keys such that decryption with incorrect keys yields honey messages---messages that 
are as likely to occur as the original message, but are hard to guess {\em without a database breach}~\cite{2014:honey-enc, 2024:bernoulli-honeywords}. We consider the event of a full database breach. HE makes it difficult for the adversary to distinguish correct from incorrect decryption, thwarting offline message recovery attacks. It can aid in breach detection by alerting when an adversary unknowingly queries the database with the honeyword. 
Moreover, password encryption schemes generally do not require locality-sensitivity.
\section{Conclusions}

We show that fuzzy extractors offer attack-agnostic post-processing defenses against model inversion attacks on ML embeddings leaked from a database breach. Our work initiates the cryptanalysis of $\ell_2$-based fuzzy extractors, showing that prior schemes are not secure when used as-is for real-world face authentication. We have proposed \prothash,
a protection scheme that enables vector comparison under $\ell_2$ distance. It offers strong security even when all persistently stored data is leaked in a breach, as evidenced in our experiments where \prothash effectively defends against all evaluated inversion attacks. Our scheme, definitions, and {theorems are generic and can be applied} to other noisy sources, such as text and audio embeddings.

\section*{Acknowledgments}
We thank NUS KISP Lab members, Divesh Aggarwal, and Tran Ngo for their insightful feedback.

\bibliographystyle{IEEEtran}
\bibliography{references}

\appendices
\section{Open-Source Artifact}
Please see \repourl~for the artifact for this paper.

\section{
Formalization of the Ideal Primitive
}
\label[app]{sec:formal-stuff}

\paragraph{\bf Notation.} Aside from symbols specified in~\Cref{tab:symbol-legend}, we use uppercase alphabets for sets, metric spaces, and random variables. We use cursive uppercase alphabets for probability distributions. Specifically, $\calU(S)$ denotes the uniform random distribution over the set $S$. 
We use {$\draw{\calD} Q$ to denote sampling from a set $Q$ according to distribution $\calD$}, or simply $\gets Q$ when the underlying distribution $\calD$ is obvious from context. 
$\X$ and $\Y$ are input and output sets from a metric space with $\dist=\ell_2$ distance, such as $\X=\R^\m$ and $\Y=\R^n$. 
The ball $B_m(\x,t)$ is the set of all vectors within distance $t$ of $\x\in\R^m$. 
We also use $B(x,t)$ to denote the ball $B_m(\x,t)$ when the omitted variable is clear from context.

\label{sec:formal-def}

\begin{definition}[Ideal primitive]
\label{def:fuzzy-ow-hash}
    An ($\X, \calW, \Y,  t, \lambda$) ideal primitive is a pair of randomized procedures, ``generate” (\gen) and ``reproduce” (\rep). \gen on input $\x\in \X$ outputs $\y\in \Y$ and a helper {$\s\in\bit^*$}. \rep takes $\x'\in \X$ and {$\s\in\bit^*$} as inputs and gives some $\y'\in \Y$ as output. (\gen, \rep) have the following properties: 
    \begin{enumerate}
    \item \label[prop]{prop:noise-tol}
    Correctness (Noise tolerance): If $\dist(\x,\x')\leq t$ and $(\y, \s) = \gen(\x)$, then $${\Pr}[\y=\rep(\x',\s)] \geq 1 - \delta.$$
    
    \item \label[prop]{prop:approx-non-inv}
    Security (Fuzzy one-way): Given $(\y, \s)$ from $\gen(\x)$ where $\x\draw{\calW} \X$, it is hard to find $\x'$ such that $\dist(\x,\x')\leq~t$. 
    In other words, for any {computationally bounded} adversary $\calA$ trying to find such $\x'$, 
    it succeeds with probability $\leq \negl(\lambda)$.

    \item \label[prop]{prop:entropy}
    Utility (Entropy sufficiency): Let $\calY$ and $\calS$ be the output distributions of $\gen$ for an input distribution $\calW$. The HILL entropy of $\calY|\calS$ is at least the min entropy of $\calW$ with entropy loss of $\tau$. 
    \end{enumerate}
\end{definition}

The probability in~\Cref{prop:noise-tol} is taken over $\x' \draw{\calW} {B}(\x,t)$ and $\x\draw{\calW} \X$. The $\delta$ term in~\Cref{prop:noise-tol} can be any value in $[0,1]$. It only affects functional correctness, such as the false positives and negatives in the authentication application. On the other hand, ``hardness" in~\Cref{prop:approx-non-inv} is cryptographic, i.e., the probability of an adversary violating~\Cref{prop:approx-non-inv} should be a negligible function in $\lambda$. {We consider a computationally bounded adversary that runs in $\poly(\lambda)$ time}. We formalize the security game of~\Cref{prop:approx-non-inv} in~\Cref{sec:ideal-sec-game}. In~\Cref{prop:entropy}, HILL entropy is the computational notion of min entropy~\cite{1999:extractor-universal-hash-hill}.

\section{\owfe Implements the Ideal Primitive}
\label[app]{app:fe-is-fowh}

We prove in~\Cref{thm:fe-is-fowh} that a computationally secure \owfe achieves all the properties required by an ideal primitive\louise{ under the computationally bounded error model~\cite[Sec.~8]{2008:fe-journal}}. 

\ThmFEisFOWH*

\subsection{Definitions and Lemmas}

The entropy notion in the utility property is formally defined in~\Cref{def:hill-entropy}, where $\comp$ denotes computational indistinguishability with respect to parameter $\epsilon$. 

\begin{definition}[HILL Entropy~{\cite{1999:extractor-universal-hash-hill}}]
\label{def:hill-entropy}
    Let $X, Z$ be random variables. HILL computational entropy $\Hhill_\epsilon(X\mid Z)$ is at least $\l$ if there exists a random variable $Y$ such that average min-entropy $\Hminavg(Y\mid Z)\geq \l$ and $(X, Z) \comp (Y,Z)$.
\end{definition}

The security games for the \fe primitive, the ideal primitive, and the one-way function are described below.

{\bf Game $\calG_0$ (FE security game):}
\begin{itemize}
        \item Challenger $\calC$ flips a bit $b \gets \calU(\bit)$.
    \begin{itemize}
        \item If $b = 1$, $\calC$ samples $x \gets \calX$, and computes $(r, p) \gets \gen(x)$.
        \item If $b = 0$, $\calC$ samples $x \gets \calX$, computes $(\_, p) \gets \gen(x)$, and samples $r \gets \calU(\bit^\l)$.
    \end{itemize}
    \item Adversary $\calA$ receives $(r, p)$ and outputs a guess $b'$.
\end{itemize}
The game $\calG_0$ outputs $1$ if $b' = b$. The advantage of $\calA$ is
\[
\adv(\calA) = \left| \Pr[\louise{b'} = 1 \mid b = 1] - \Pr[\louise{b'} = 1 \mid b = 0] \right|.
\]

{\bf Game $\calG_1$ (Ideal primitive security game):}
\label{sec:ideal-sec-game}
\begin{itemize}
    \item Challenger $\calC$ samples $x \gets \calW$ and computes $(r, p) \gets \gen(x)$.
    \item Adversary $\calA$ receives $(r, p)$ and outputs a guess $x'$.
\end{itemize}
The game $\calG_1$ outputs $1$ if $d(x, x') \leq t$,
\louise{so the success probability is 
}
\[
\Pr[\calG_1(\calA)=1] = \louise{\Pr\left[\dist(x, \calA(r,p)) \leq t\right].}
\]
\louise{The ideal primitive's \Cref{prop:approx-non-inv} holds if $\Pr[\calG_1(\calA)=1] \leq \negl(\lambda)$ for every PPT $\calA$.}

\louise{{\bf Game $\calG_2$ (One-wayness of $H_k$, \Cref{def:ow-hash}):}
\begin{itemize}
    \item Challenger $\calC$ samples $k \gets K$ and $u \gets \calU(\bit^\l)$.
    \item Adversary $\calA$ receives $(k, u)$ and outputs guess $z$.
\end{itemize}
$\calG_2$ outputs $1$ if $H_k(z) = u$. 
In other words, $\{H_k\}$ is $\epsilon_{\mathsf{OW}}$-one-way if $\Pr[\calG_2(\calA)=1]\leq \epsilon_{\mathsf{OW}}$ for every PPT $\calA$.
}

\louise{We prove in~\Cref{thm:fe-security-reduction} that an adversary for $\calG_1$ yields an adversary for $\calG_0$ and an adversary for $\calG_2$.}

\begin{lemma}
\label{thm:fe-security-reduction}
\louise{Let $(\gen, \rep)$ be an $\epsilon_{\fe}$-secure \owfe whose hash family $\{H_k\}_{k \in K}$ is $\epsilon_{\mathsf{OW}}$-one-way. Then every PPT adversary $\calA_1$ for $\calG_1$, with output $x' \gets \calA_1(r, p)$, satisfies}
\[
\louise{
\Pr\left[\dist(x, x') \leq t \text{ and } \rep(x', p) = r\right] \leq \epsilon_{\fe}+\epsilon_{\mathsf{OW}}.}
\]
\end{lemma}

\begin{proof}
\louise{Let $\calA_1$ be a PPT adversary for $\calG_1$. By~\Cref{def:owfe}, the helper has the form $p = (\mathsf{aux}, k)$, where $\mathsf{aux}$ is computed by $\mathsf{Gen}_0(x)$ and $k \gets K$ is sampled independently, and $\rep(x', p) = H_k(\mathsf{Rep}_0(x', \mathsf{aux}))$. }We construct $\calA_0$ for game $\calG_0$ as follows: \louise{On input $(r, p)$,}
\begin{itemize}
    \item Run $x' \gets \calA_1(r, p)$. 
    \item Compute $r' \gets \rep(x', p)$. 
    \item If $r=r'$, return $b':=1$. Else return $b':=0$.
\end{itemize}
\louise{For each possible value $a$ of $\mathsf{aux}$, let $\calA_2^{a}$ be the following adversary for game $\calG_2$, with $a$ built into its description: On input $(k, u)$,}
\begin{itemize}
    \item \louise{Run $x' \gets \calA_1(u, (a, k))$.}
    \item \louise{Return $z := \mathsf{Rep}_0(x', a)$.}
\end{itemize}
\louise{
We know by assumption that $(\gen,\rep)$ satisfies $\adv(\calA_0) \leq \epsilon_{\fe}$ and $H_k$ satisfies $\Pr[\calG_2(\calA_2^{a})=1] \leq \epsilon_{\mathsf{OW}}$ for every $a$.
}

Now we analyze advantage of $\calA_0$. 

For the $b=1$ case, the $(r,p)$ are genuine pairs, hence
\begin{align*}
\louise{\Pr[\calA_0 \text{ outputs } 1 | b = 1]} 
&=\Pr[r = r']\\
&\geq \louise{\Pr[\dist(x, x') \leq t \text{ and } \rep(x', p)=r]}.
\end{align*}

\louise{When $b = 0$, the key $r$ is a fresh uniform string, independent of $x$ and of $p = (\mathsf{aux}, k)$, and $\mathsf{aux}$ is computed without $k$. So $(k, r)$ is distributed exactly as a $\calG_2$ challenge $(k, u)$, independent of $\mathsf{aux}$. $\calA_0$ outputs $1$ only if $H_k(\mathsf{Rep}_0(x', \mathsf{aux})) = r$, which is exactly the event that $\calA_2^{\mathsf{aux}}$ wins $\calG_2$ on challenge $(k, r)$. Averaging over $\mathsf{aux}$,}
\begin{align*}
\louise{\Pr[\calA_0 \text{ outputs } 1 | b = 0] = \mathbb{E}_{\mathsf{aux}}\left[\Pr[\calG_2(\calA_2^{\mathsf{aux}})=1]\right] \leq \epsilon_{\mathsf{OW}}.}
\end{align*}

\louise{Combining the two cases,
\[
\epsilon_{\fe} \geq \adv(\calA_0) \geq \Pr\left[\dist(x, x') \leq t \text{ and } \rep(x', p) = r\right] - \epsilon_{\mathsf{OW}},
\]
which is the lemma.
}
\end{proof}

\subsection{Theorem Proof}
\begin{proof}[Proof of~\Cref{thm:fe-is-fowh}]

{\bf Correctness proof.}
\louise{It follows directly from the \fe correctness property in~\Cref{def:fe-error}. 
} 

{\bf Utility proof. }
Since $\Hmin(\calU(\bit^\l) \mid \calP) = \Hmin(\calU(\bit^\l)) = \l$, and $(\calR, \calP)\approx_{\epsilon_{\fe}} (\calU(\bit^\l), \calP)$ by \fe security, then $\Hhill(\calR \mid P)\geq \l$ by \Cref{def:hill-entropy}. Hence the utility property~\Cref{prop:entropy} of the ideal primitive holds with entropy loss at most $\max\{0,\Hmin(\calX)-\ell\}$.

{\bf Security proof.} 
\louise{Let $\calA$ be an efficient adversary for $\calG_1$ with output $x' \gets \calA(r, p)$, and let $r' \gets \rep(x', p)$. Splitting the winning event on whether $x'$ reproduces the key,
\begin{align*}
\Pr[\calG_1(\calA)=1] &= \Pr[\dist(x, x') \leq t \text{ and } r' = r]\\
&\quad + \Pr[\dist(x, x') \leq t \text{ and } r' \neq r]\\
&\leq (\epsilon_{\fe}+\epsilon_{\mathsf{OW}}) + \alpha.
\end{align*}
The first term is bounded by~\Cref{thm:fe-security-reduction}, and the second by the definition of $\alpha$ in~\Cref{thm:fe-is-fowh}. The sum is at most $\negl(\lambda)$ by assumption, hence~\Cref{prop:approx-non-inv} holds.} 

\end{proof}

\louise{\textbf{Remark.} The error $\delta$ of~\Cref{def:fe-error} cannot replace the $\alpha$ assumption in~\Cref{thm:fe-is-fowh}: $\delta$ only covers errors from an $x'$ chosen independently of the helper~\cite{2020:lwe-fe,2021:sample-then-lock}, whereas the adversary of $\calG_1$ can choose $x'$ after seeing $(r, p)$.}

\subsection{\louise{FE Security Does Not Imply Ideal-Primitive Security}}
\label[app]{app:fe-counterexample}

\louise{We give an \fe example scheme with perfect correctness and perfect security whose key nonetheless reveals the input. This is referred to in~\Cref{subsec:fe} as the rounding scheme.}

\louise{\emph{The rounding scheme.} Fix $\l \geq 1$. Let $\X = \{4j + b : j = 0, \ldots, 2^\l - 1,\ b \in \{0, 1\}\} \subset \mathbb{Z}$ with $\ell_1$ distance $\dist(\x, \x') = |\x - \x'|$. Let $t = 1$, and let $\calW$ be uniform over $\X$.}
\begin{itemize}
    \item \louise{$\gen(\x)$: Output the key $r = \lfloor \x/4 \rfloor$, written as an $\l$-bit string, and the helper $p = 0^\l$.}
    \item \louise{$\rep(\x', p)$: Output $\lfloor \x'/4 \rfloor$.}
\end{itemize}
\louise{For example, with $\l = 2$, the inputs in $\{0,1\}, \{4,5\}, \{8,9\}, \{12,13\}$ map to the keys $00, 01, 10, 11$, respectively. The helper is always $00$.}

\begin{lemma}
\label{lem:fe-counterexample}
\louise{The scheme above is an ($\X, \calW, \l, 1, 0$)-\fe with error $\delta = 0$.}
\end{lemma}

\begin{proof}
\louise{\emph{Correctness.} Two inputs with different $j$ are at distance at least $3$ (e.g., $4j+1$ and $4j+4$), so $\dist(\x, \x') \leq 1$ implies that $\x$ and $\x'$ share the same $j$. $\rep(\x', p) = j = r$ for every such pair, so this scheme achieves perfect correctness with $\delta = 0$.}

\louise{\emph{Security.} Each key value $r$ corresponds to exactly two inputs, $4r$ and $4r + 1$, so $r$ is uniform over $\bit^\l$, and $p = 0^\l$ is constant. Hence $(r, p)$ is distributed exactly as $(\calU(\bit^\l), 0^\l)$. This means perfect security of $\epsilon_{\fe} = 0$, even against unbounded distinguishers.}
\end{proof}

\louise{\emph{The scheme is not fuzzy one-way.} Consider an adversary that outputs $\x' = 4r$ when given $(r,p)$.
Since $\x \in \{4r, 4r+1\}$, the output satisfies $\dist(\x, \x') \leq 1$, so the adversary wins $\calG_1$ with probability $1$, and even recovers $\x$ exactly with probability $1/2$. Even though the key appears uniform as required by \fe security, it is actually just the input in binary format with one bit of information lost. Thus, all the remaining information of the input is leaked in the clear.
Choosing a larger $\l$
does not help mitigate this issue.}

\section{Entropy Estimation for our scheme}
\label[app]{sec:entropy-est-soft}
\label[app]{app:entropy-est}

{We present the baseline and elimination-based enumeration algorithms in~\Cref{alg:basic-attacker} and~\Cref{alg:elimination-attacker}, respectively}. We then adapt them to target \fe schemes that use soft-thresholding in verification, such as ours, shown in~\Cref{alg:basic-soft-attacker} and~\Cref{alg:elimination-soft-attacker}. 

{\paragraph{Baseline Enumeration.}
\algoEntropyBasic
In~\Cref{alg:basic-attacker}, the adversary enumerates $x'$ from the sample space $\X$ and computes $\repPartial(x',p)$. This results in an empirical distribution of $\repPartial(X,p)$, serving as the attacker's guessing distribution $\calR'_p$.

\paragraph{Elimination-Based Enumeration.}
\algoEntropyEliminate
In~\Cref{alg:elimination-attacker}, the attacker
groups outputs from same-identity inputs into batches, and records, for each batch, the majority value together with its vote count in the histogram. Batches that fail to converge to a majority are discarded. Compared with~\Cref{alg:basic-attacker},  this yields a lower-entropy empirical distribution by eliminating implausible outputs and reducing the distribution's support.
}

\paragraph{Baseline Enumeration with Soft Threshold.}
\algoEntropyBasicSoft
For each batch of samples corresponding to a single identity, the~\Cref{alg:basic-soft-attacker} attacker \louise{runs \rep on every sample and compares pairwise whether they are accepted by \verify, i.e. they share at least $N_t$ block hashes. 
It greedily picks the output $r^\star$ with the largest set of accepted outputs in the batch. 
Every output in its accepted set $\mathsf{Acc}(r^\star)$ is assigned to the cluster of $r^\star$, so the histogram count of $r^\star$ is incremented by the size of this set. The process is repeated on
the remaining outputs until every output in the batch has been assigned to a cluster. The resulting histogram induces the attacker's guessing distribution $\calR'_p$.}

\paragraph{Elimination-Based Enumeration with Soft Threshold.}
\algoEntropyEliminateSoft
The~\Cref{alg:elimination-soft-attacker} attack is similar to~\Cref{alg:basic-soft-attacker},
except for each batch, it only accounts for the \louise{output $r^\star$ whose accepted set contains at least half of the outputs in the batch}. If no such \louise{output} exists, the entire batch is discarded. Eliminating batches with widespread and inconsistent \louise{outputs} induces a lower-entropy 
guessing distribution $\calR'_p$.
 
\section{Parameter Selection}
\label[app]{app:cryptanalysis-results}
We first split the CelebA dataset, which contains 10177 identities, into a train set with 9177 ids and a test set with 1000 ids. The threshold for the unprotected scheme and hyperparameters for the protected schemes are selected using {1000 genuine pairs of images from the same person, and 1000 pairs of images from different people, all sampled from the set of images belonging to the train ids}. The schemes' performance in~\Cref{tab:tpr-fpr} is reported on the CelebA test split along with other unseen datasets---CASIA-Webface and LFW.

For \facialfe schemes, we pick the scale which gives us the best balanced accuracy, i.e. (TPR+1-FPR)/2, with the constraint that authentication is successful only if all 21 blocks of the stored and reproduced $r$ match exactly. For $q=1024$, the scales obtained for Leech-based and $E_8$-based \facialfe are $1461.76$ and $1/7.25$ respectively.

For \prothash, we relax the constraint that all blocks of $r$ should match. Then, we pick the best scale and block threshold ($N_t$) values that give the highest balanced accuracy. This gives us $N_t = 4$ and scale = $144.64$.

For \randomfe, the LLL reduction algorithm is a computational bottleneck, hence we project the $512$-dimensional embedding to $m = 180$ dimensions and use $n = 60$ dimensions for the random vector $b$. Similar to \facialfe, the scale parameter is chosen such that the balanced accuracy is maximized on the train set. The scale for \randomfe is $0.0014q$, where $q = 130003$.

\section{Ablation studies}
\label[app]{app:ablation}

\paragraph{Model checkpoints.}
\imgPipeCkptASR
{\Cref{fig:pipe-ckpt-asr} shows \attackname \asrrep and \asremb as the number of training steps increases. ASR for unprotected and \facialfeE schemes reaches near 100\% with 25000 steps for \attackname attack. Furthermore, the increasing trend for Leech-based \facialfe schemes indicates that ASR could increase further if trained for more steps. On the contrary, the stably low ASR (near random guessing) for \prothash suggests that 
additional training is unlikely to improve the attack success significantly. }

\paragraph{Majority voting.}
{\Cref{fig:majority-vote} shows the improvement in TPR from $85.1\%$ to $93.4\%$ and FPR from $4.6\%$ to $1.3\%$ of \prothash as the number of images used for majority voting increases from $1$ to $25$. \louise{We use the $1000$ CelebA identities with the most images. For each identity, we enroll one image, and draw authentication images from its remaining images for genuine attempts and from the images of the other identities for impostor attempts. In a face authentication system, the multiple images captured at authentication time can be evaluated in parallel to reduce overhead.}
\imgMajorityVote

\section{Security Proof for \prothash~(\Cref{algo:l2fe-hash-decode})}
\label[app]{sec:new-scheme-sec-proof}

\subsection{Proof of \fe Security}

Before proving~\Cref{thm:new-scheme-sec} \louise{that \prothash satisfies \fe security}, we first establish~\Cref{lem:leftover-hash}, which follows from~\cite{2008:fe-journal} Lemma 2.4. 

\begin{lemma}[General Leftover Hash Lemma]
\label{lem:leftover-hash}
Let $\{H_k \colon \calL_q \to \bit^\ell\}_{k\in K}$ be a family of universal hash functions. Let $\calD$ denote some distribution over $\calL_q$, and $\calD'$ denote arbitrary side information correlated to $\calD$. Let $h:= \Hmin(\calD | \calD')$. Let $\calK$ denote the uniform distribution over $K$.  Then
\[
\sd((H_\calK(\calD), \calK, \calD'), (\calU_\ell, \calK, \calD')) \leq \frac{1}{2}\sqrt{2^{-h}2^\ell}
\]
where $\calU_\ell = \calU(\bit^\ell)$. In other words, for $\ell \leq h - 2\log(1/\epsilon)+2$, we have
\[
\sd((H_\calK(\calD), \calK, \calD'), (\calU_\ell, \calK, \calD')) \leq \epsilon
\]
\end{lemma}

\begin{proof}[\bf Proof of~\Cref{thm:new-scheme-sec}]
For $j=0,\ldots,N_b$, define the hybrid distribution $\mathsf{H}_j$
\[
\mathsf{H}_j :=
\begin{aligned}
&(\calK_{1},H_{\calK_{1}}(\calD_{1}),
 \ldots,\calK_{j},H_{\calK_{j}}(\calD_{j}),\\
&\calK_{j+1},\calU_{j+1},
 \ldots,
 \calK_{N_b},\calU_{N_b})
\end{aligned}
\]
where $U_1,\ldots,U_{N_b}$ are independent uniform random variables over
$\bit^{\ell'}$, independent of all other variables.
Thus $\mathsf{H}_{N_b}$ is the real distribution $(\calP,\calR)$, and
$\mathsf{H}_{0}$ is the ideal distribution $(\calP,\calU_{\ell'}^{N_b}) = (\calP,\calU_\ell)$.

Fix $j\in[N_b]$.
Conditioned on $\calD_1,\ldots,\calD_{j-1}$, we have 
\[
    \Hmin(\calD_j\mid \calD_1,\ldots,\calD_{j-1})\ge h.
\]
Then by applying~\Cref{lem:leftover-hash}, 
\[
\begin{aligned}
&\sd\left(
    (H_{\calK_j}(\calD_j),\calK_j,\calD_{j-1},\ldots,\calD_{1}),
\right.\\
&\quad\left.
    (U_j,\calK_j,\calD_{j-1},\ldots,\calD_{1})
\right)
\le \epsilon .
\end{aligned}
\]
To complete the hybrid, we can include independent randomness $\calK_1, ..., \calK_{j-1},\calK_{j+1}, ..., \calK_{N_b}$ and $\calU_{j+1}, ..., \calU_{N_b}$, then post-process $\calD_i$ into $H_{\calK_i}(\calD_i)$ for all $i< j$ without increasing the statistical distance. 
Hence we bound the distance between consecutive hybrids,
\[
    \sd(\mathsf{H}_{j},\mathsf{H}_{j-1})\le \epsilon
\]

By the triangle inequality,
\[
    \sd(\mathsf{H}_{N_b},\mathsf{H}_{0})
    \le
    \sum_{j=1}^{N_b}\sd(\mathsf{H}_{j},\mathsf{H}_{j-1})
    \le
    N_b\epsilon .
\]

Therefore,
$
    \sd\left((\calR,\calP),(\calU_\ell,\calP)\right)
    \le N_b\epsilon.
$
\end{proof}

\subsection{\louise{Proof of Fuzzy One-wayness}}
\louise{
We now prove the security property of the ideal primitive (\Cref{prop:approx-non-inv}) for \prothash. 
As in~\Cref{sec:proposed-prot-scheme}, we model the keyed hash as a random oracle $\calO$ that the adversary can query, i.e., $H_k(\cdot) = \calO(k, \cdot)$}.

\begin{thm}
\label{thm:snapper-fow}
\louise{Let $H_k(\cdot) = \calO(k, \cdot)$ for a random oracle $\calO$, with keys drawn uniformly from $K$, and let $h$ be as in~\Cref{thm:new-scheme-sec}. Every adversary that makes at most $Q$ queries to $\calO$ and outputs $\x'$ given $(r, p)$ from \prothash satisfies
\[
\Pr[\dist(\x,\x')\leq t] \leq 2^{-\Hfuzz(\calW)} + Q\cdot 2^{-h} + N_b^2/|K|,
\]
where $2^{-\Hfuzz(\calW)} := \max_{\x'}\Pr_{\x\gets\calW}[\dist(\x,\x')\leq t]$ defines the fuzzy min-entropy of the input distribution $\calW$~\cite{2020:fe-impossible}.}
\end{thm}

\louise{Each term corresponds to one way for the adversary to win:
\begin{itemize}
    \item $2^{-\Hfuzz(\calW)}$: it guesses $\x'$ without using $(r, p)$; no scheme can avoid this term.
    \item $Q\cdot 2^{-h}$: it guesses a decoded block of $\x$ and checks the guess against the stored hash with an oracle query. Each of the $Q$ guesses is correct with probability at most $2^{-h}$.
    \item $N_b^2/|K|$: two blocks are hashed with the same key, in which case equal hashes reveal equal blocks.
\end{itemize}}

\begin{proof}[roof of~\Cref{thm:snapper-fow}]
\louise{Recall that \gen decodes $\x$ into blocks $d = (d_1, \ldots, d_{N_b})$, samples keys $k_1, \ldots, k_{N_b} \gets K$, and outputs $p = (k_1, \ldots, k_{N_b})$ and $r = (r_1, \ldots, r_{N_b})$ with $r_i = \calO(k_i, d_i)$. Let $\calA$ be an adversary for $\calG_1$ (\Cref{app:fe-is-fowh}) that makes at most $Q$ queries to $\calO$. We compare $\calG_1$ with the following game, in which $\calA$ receives random strings in place of the hashes.}

\louise{{\bf Game $\calG_1'$ (random strings in place of hashes):}
\begin{itemize}
    \item Challenger $\calC$ samples $x \gets \calW$, computes $(r, p) \gets \gen(x)$, and samples $u \gets \calU(\bit^\l)$.
    \item Adversary $\calA$ receives $(u, p)$, queries $\calO$ at most $Q$ times, and outputs a guess $x'$.
\end{itemize}
The game $\calG_1'$ outputs $1$ if $\dist(x, x') \leq t$.}

\louise{We define two events:
\begin{itemize}
    \item $\mathsf{Coll}$: two keys are equal, i.e., $k_i = k_j$ for some $i \neq j$.
    \item $\mathsf{Bad}$: $\calA$ queries $\calO$ at $(k_i, d_i)$ for some block $i$.
\end{itemize}}

\louise{\paragraph{Relating the two games.}
Suppose $\mathsf{Coll}$ does not occur. Then the points $(k_i, d_i)$ are distinct, so the hashes $r_i = \calO(k_i, d_i)$ are uniform, independent of each other, and independent of the value of $\calO$ at every other point. Hence $\calG_1$ is distributed exactly as $\calG_1'$ with $\calO$ reprogrammed to output $u_i$ at $(k_i, d_i)$, and $\calA$ can only notice the difference if it queries one of these points, i.e., if $\mathsf{Bad}$ occurs. Therefore
\begin{align*}
\Pr[\calG_1(\calA)=1] \leq{}& \Pr[\calG_1'(\calA)=1] + \Pr[\mathsf{Coll} \text{ occurs}]\\
&+ \Pr[\mathsf{Bad} \text{ occurs and } \mathsf{Coll} \text{ doesn't occur}].
\end{align*}
Now we bound the three terms.}

\louise{\paragraph{Bounding first term.}
In $\calG_1'$, the adversary only sees $u$, $p$ and the answers of $\calO$. All of these are independent of $x$, so its output $x'$ is independent of $x$ as well. Hence, whatever $x'$ the adversary outputs, $x$ is still a fresh sample from $\calW$, and it lands within distance $t$ of $x'$ with probability at most the maximum over all points:
\[
\Pr[\calG_1'(\calA)=1] \leq \max_{\x'} \Pr\nolimits_{\x \gets \calW}[\dist(\x, \x') \leq t] = 2^{-\Hfuzz(\calW)}.
\]}

\louise{\paragraph{Bounding second term.}
The keys are independent and uniform over $K$, so any two of them are equal with probability $1/|K|$. By a union bound over the pairs of blocks,
\[
\Pr[\mathsf{Coll} \text{ occurs}] \leq \binom{N_b}{2}/|K| \leq N_b^2/|K|.
\]}

\louise{\paragraph{Bounding third term.}
For the same reason as above, every query $(k', d')$ of $\calA$ in $\calG_1'$ is independent of the decoded blocks $d$ of $x$. When the keys are distinct, $k'$ equals at most one key $k_i$, and the query then hits $(k_i, d_i)$ only if $d' = d_i$. This happens with probability at most
\[
\max_{d'} \Pr[\calD_i = d'] = 2^{-\Hmin(\calD_i)} \leq 2^{-h},
\]
where $\Hmin(\calD_i) \geq \Hmin(\calD_i \mid \calD_1, \ldots, \calD_{i-1}) \geq h$
By a union bound over the $Q$ queries,
\[
\Pr[\mathsf{Bad} \text{ occurs }\text{ and } \mathsf{Coll} \text{ doesn't occur}] \leq Q \cdot 2^{-h}.
\]
Summing the three bounds proves the claim.}
\end{proof}

\louise{
It follows that~\Cref{prop:approx-non-inv} holds for \prothash against adversaries that make polynomially many queries when $\Hfuzz(\calW)$ and $h$ are $\omega(\log\lambda)$ and $|K|$ is superpolynomial in $\lambda$, e.g., for $\lambda$-bit keys.} 

\subsection{\louise{\prothash Is an Ideal Primitive}}
\label[app]{sec:prothash-is-ideal}
\louise{The three properties of the ideal primitive hold for \prothash as follows:
\begin{itemize}
    \item Correctness (\Cref{prop:noise-tol}) follows from \fe correctness, similar to the proof of~\Cref{thm:fe-is-fowh}.
    \item Security (\Cref{prop:approx-non-inv}) follows from~\Cref{thm:snapper-fow}.
    \item Utility (\Cref{prop:entropy}) follows from \fe security shown in~\Cref{thm:new-scheme-sec}, similar to the proof of~\Cref{thm:fe-is-fowh}.
\end{itemize}
Although \Cref{thm:new-scheme-sec} is stated for a universal hash, it also holds for a random oracle hash. The proof of~\Cref{thm:snapper-fow} shows that an adversary making at most $Q$ queries can tell $r$ apart from a uniform string only if $\mathsf{Bad}$ or $\mathsf{Coll}$ occurs, which has probability at most $Q\cdot 2^{-h}+N_b^2/|K|$. So $r$ is still indistinguishable from uniform given $p$, which is the conclusion of~\Cref{thm:new-scheme-sec}. Hence all three properties hold for \prothash instantiated with a random oracle hash.}

\section{Prior Inversion Attacks}
\label[app]{app:prior-attacks}

In~\Cref{sec:pipe-asr}, we have shown that \attackname has a high ASR against \facialfe schemes. Are these schemes effective against prior attacks to begin with?
To answer this, we evaluate $3$ state-of-the-art inversion attacks on face recognition models: \bob~\cite{2024:bob}, KEDMI~\cite{2021:ked-mi}, GMI~\cite{2020:gmi}. \bob is the state-of-the-art white-box CNN-based embedding inversion attack.
KEDMI and GMI are state-of-the-art black-box model inversion attacks\footnote{The core of these attacks only requires black-box access to the target model, and a separate inversion model trained on the public dataset.}
~\cite{2024:midre}. They both use GANs as the inversion model and have the best-known performance among black-box model inversion attacks~\cite{2022:bido}. 

\tableLPIPS
\tableFID
\paragraph{Evaluation Setup.} KEDMI and GMI are designed for closed-set face classification systems, so we append a classification layer to the target model and provide its output probabilities to these attacks. The classification layer is trained and tested on disjoint CelebA subsets containing the same 1,000 identities, while another disjoint subset is used for training the inversion model. Because the protection schemes are not optimized for classification, classifier accuracy ranges from 0.17\% to 87.13\% depending on the scheme. KEDMI and GMI are evaluated on identities seen during classifier training. Since they reconstruct an image for a target identity, not a specific enrolled image, we consider a reconstruction to succeed if its embedding is within $t$ of any of 10 training images of that identity. This is looser than \asremb, which compares against a single original image, so the closed-set and open-set results are not directly comparable.
\attackname and Bob are evaluated in the more challenging open-set setting on unseen identities, hence Bob is the most directly comparable baseline for \attackname.

\paragraph{Realistic images.}
As seen in~\Cref{tab:LPIPS,tab:FID}, \attackname achieves better visual similarity indicated by lower LPIPS~\cite{2018:LPIPS} and FID scores~\cite{2017:fid}. 
This further confirms that \attackname creates perceptually closer reconstructions to the original images.

\end{document}